\documentclass[letterpaper]{article} 
\usepackage{aaai2026}  
\usepackage{times}  
\usepackage{helvet}  
\usepackage{courier}  
\usepackage[hyphens]{url}  
\usepackage{graphicx} 
\usepackage{natbib}  
\usepackage{caption} 
\usepackage{amsmath}
\usepackage{amssymb}
\usepackage{algorithm}
\usepackage{algpseudocode}
\usepackage{graphicx}
\usepackage{xcolor}
\usepackage{tabularx} 
\usepackage{subfig} 
\usepackage{booktabs}
\usepackage{multirow}
\usepackage{array}
\usepackage{colortbl}
\usepackage{graphicx}
\definecolor{rob0}{RGB}{242,246,250}
\definecolor{rob1}{RGB}{228,236,245}
\definecolor{rob2}{RGB}{208,223,239}
\definecolor{rob3}{RGB}{183,209,231}
\definecolor{rob4}{RGB}{153,190,220}
\definecolor{rob5}{RGB}{114,166,206}
\definecolor{rob6}{RGB}{72,138,191}

\definecolor{rec0}{RGB}{244,248,242}
\definecolor{rec1}{RGB}{232,241,227}
\definecolor{rec2}{RGB}{216,234,208}
\definecolor{rec3}{RGB}{194,224,183}
\definecolor{rec4}{RGB}{165,211,151}
\definecolor{rec5}{RGB}{128,194,112}
\definecolor{rec6}{RGB}{86,174,79}
\definecolor{rec7}{RGB}{49,153,62}

\newcommand{\robzero}[1]{\cellcolor{rob0}{#1}}
\newcommand{\robone}[1]{\cellcolor{rob1}{#1}}
\newcommand{\robtwo}[1]{\cellcolor{rob2}{#1}}
\newcommand{\robthree}[1]{\cellcolor{rob3}{#1}}
\newcommand{\robfour}[1]{\cellcolor{rob4}{#1}}
\newcommand{\robfive}[1]{\cellcolor{rob5}{\textcolor{white}{#1}}}
\newcommand{\robsix}[1]{\cellcolor{rob6}{\textcolor{white}{#1}}}

\newcommand{\reczero}[1]{\cellcolor{rec0}{#1}}
\newcommand{\recone}[1]{\cellcolor{rec1}{#1}}
\newcommand{\rectwo}[1]{\cellcolor{rec2}{#1}}
\newcommand{\recthree}[1]{\cellcolor{rec3}{#1}}
\newcommand{\recfour}[1]{\cellcolor{rec4}{#1}}
\newcommand{\recfive}[1]{\cellcolor{rec5}{\textcolor{white}{#1}}}
\newcommand{\recsix}[1]{\cellcolor{rec6}{\textcolor{white}{#1}}}
\newcommand{\recseven}[1]{\cellcolor{rec7}{\textcolor{white}{#1}}}
\newcommand{\answerYes}[1]{\textcolor{blue}{#1}} 
\newcommand{\answerNo}[1]{\textcolor{teal}{#1}} 
\newcommand{\answerNA}[1]{\textcolor{gray}{#1}} 
 
\usepackage[most]{tcolorbox}

\definecolor{promptbg}{HTML}{F8F9FB}
\definecolor{promptframe}{HTML}{D8DEE9}
\definecolor{prompttitle}{HTML}{EEF2F7}
\definecolor{prompttext}{HTML}{24292F}

\newtcblisting{promptbox}[1]{
  enhanced,
  breakable,
  listing only,
  colback=promptbg,
  colframe=promptframe,
  boxrule=0.45pt,
  arc=1.5mm,
  left=1.5mm,
  right=1.5mm,
  top=1mm,
  bottom=1mm,
  title={#1},
  fonttitle=\bfseries\small,
  coltitle=black,
  colbacktitle=prompttitle,
  boxed title style={
    colback=prompttitle,
    colframe=promptframe,
    boxrule=0.45pt,
    arc=1mm
  },
  listing options={
    basicstyle=\ttfamily\footnotesize\color{prompttext},
    breaklines=true,
    columns=fullflexible,
    keepspaces=true,
    showstringspaces=false,
    numbers=left,
    numberstyle=\tiny\color{gray},
    xleftmargin=1.5em
  }
}
\usepackage{newfloat}
\usepackage{listings}
\DeclareCaptionStyle{ruled}{labelfont=normalfont,labelsep=colon,strut=off} 
\floatstyle{ruled}
\newfloat{listing}{tb}{lst}{}
\floatname{listing}{Listing}
\title{You Can't Fool Us: Understanding the Resilience of LLM-driven Agent Communities to Misinformation}
\author{
    Chichen Lin\textsuperscript{\rm 1},
    Yijie Jin\textsuperscript{\rm 2},
    Kangbo Hu\textsuperscript{\rm 2},
    Weijian Fan\textsuperscript{\rm 3},
    Han Xiao\textsuperscript{\rm 4},
    Yongbin Wang\textsuperscript{\rm 1},
    Zhihui Ying\textsuperscript{\rm 1},
    Zhanzhan Zhao\textsuperscript{\rm 5}\thanks{Corresponding author.}
}

\affiliations{
    \textsuperscript{\rm 1}State Key Laboratory of Media Convergence and Communication, Communication University of China, Beijing, China\\
    \textsuperscript{\rm 2}School of Computer and Cyber Sciences, Communication University of China, Beijing, China\\
    \textsuperscript{\rm 3}School of Data Science and Intelligent Media, Communication University of China, Beijing, China\\
    \textsuperscript{\rm 4}School of Information Communication, Communication University of China, Beijing, China\\
    \textsuperscript{\rm 5}School of Humanities Social Science, School of Artificial Intelligence, Chinese University of Hong Kong, Shenzhen, China\\
    cuc\_lcc@cuc.edu.cn,
    202312033012@mails.cuc.edu.cn,
    hkbyyds@cuc.edu.cn,
    wjfan@cuc.edu.cn,
    hanxiao@cuc.edu.cn,
    ybwang@cuc.edu.cn,
    yingzhihui@cuc.edu.cn,
    zhanzhanzhao@cuhk.edu.cn
}

\usepackage{bibentry}

\begin{document}

\maketitle

\begin{abstract}
    Misinformation resilience is a dynamic community process: communities differ not only in whether they initially trust false claims, but also in how they recover through interaction, questioning, correction, and support withdrawal. We study this process with an LLM-based agent simulation that constructs synthetic communities along two theoretically motivated dimensions: Actively Open-minded Thinking (AOT), which captures evidence-seeking and willingness to revise beliefs, and Political Ideology (PI), which captures identity-based interpretation of contested claims. These two traits allow us to examine how evidence-oriented reasoning and ideological alignment jointly shape community responses to credible misinformation shocks. Across systematically varied AOT–PI communities, we find that higher AOT improves both resistance to misinformation uptake and recovery after trust peaks. PI shapes the recovery pathway: ideologically moderate communities recover more reliably, while polarized communities retain more residual support. Stance-level analysis shows that resilience depends on whether agents move from questioning a claim to denying or correcting it and withdrawing prior support. Intervention experiments further show that persuasion and fact checking better support post-peak correction, whereas accuracy prompts mainly induce early caution and source warnings have weaker effects. Together, this work provides a mechanism-level account of community misinformation resilience, showing how psychological composition and intervention design shape whether communities move from misinformation exposure toward correction or persistent support.
\end{abstract}


\begin{figure}[t]
    \centering
    \vspace{-4pt}
    \includegraphics[
        width=0.95\linewidth,
    ]{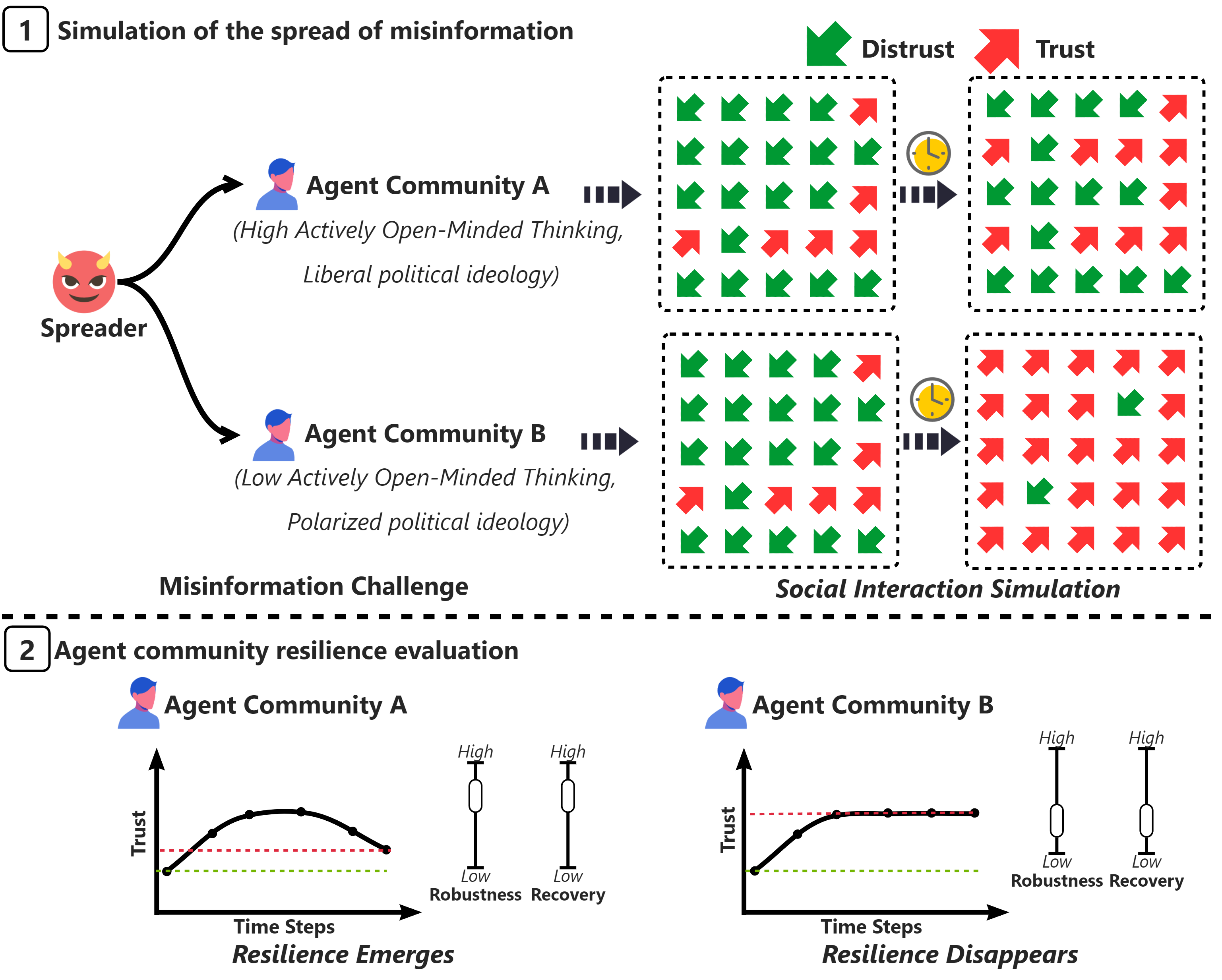}
    \vspace{-4pt}
    \caption{
    The same misinformation can drive different communities toward opposite collective outcomes. Communities with distinct psychological compositions respond differently to an identical misinformation shock: one community gradually restores distrust through social interaction, while another shifts toward persistent trust and loses resilience.
    }
    \label{fig:Resilience}
    \vspace{-8pt}
\end{figure}

\section{Introduction}\label{Introduction}

Online misinformation is not only a problem of whether individuals believe false claims, but also a problem of whether communities can recover from them. In real-world information ecosystems, the same misinformation exposure can lead to sharply different community-level outcomes: some communities gradually return to baseline skepticism, while others stabilize into persistent support for false narratives. As illustrated in Figure~\ref{fig:Resilience}, these divergent trajectories suggest that resilience is not solely a property of information content, but an emergent outcome of social interaction, psychological composition, and stance revision over time.

Existing work on misinformation has primarily focused on individual susceptibility and static predictors of belief, including reasoning style, prior knowledge, source credibility, and political ideology \citep{pennycook2019lazy,bago2020fake,pennycook2021psychology,ecker2022psychological,sultan2024susceptibility}. In parallel, intervention research has shown that fact checking, warning labels, inoculation, accuracy prompts, and corrective messages can reduce belief in false claims or discourage misinformation sharing \citep{nyhan2010corrections,chan2017debunking,porter2021global,roozenbeek2019fake,pennycook2021shifting,roozenbeek2021accuracy,roozenbeek2022psychological,bode2024user,altay2025small}. Yet these lines of work leave open a central collective question: how do heterogeneous communities recover, or fail to recover, after misinformation has already entered the social system? In particular, we know less about how individual differences combine through repeated interaction to produce community-level recovery trajectories, residual support, or persistent polarization.

Recent advances in large language model agents provide a foundation for studying such dynamics in controlled environments. LLM agents can represent stable psychological profiles, maintain memory, and engage in multi-round natural language interaction, enabling simulation of complex social systems \citep{park2023generative,park2024generative,mou2024individual}. Prior work has explored LLM-based simulations of information diffusion, persuasion, and moderation dynamics \citep{li2024newsdiffusion,liu2025mosaic,liu2025stepwise,borah2026persuasion}. Yet existing frameworks remain largely propagation-centric: they often focus on how information spreads or how engagement changes, while offering a limited understanding of resilience as an emergent process arising from psychological heterogeneity, interpersonal interaction, and stance evolution within communities.

To address this gap, we move from individual-level predictors to the population-level composition of misinformation resilience. Building on prior work that links misinformation susceptibility to cognitive reasoning and social identity \citep{osmundsen2021partisan,roozenbeek2022susceptibility,murphy2019false,biddlestone2025norm,sultan2024susceptibility}, we focus on two corresponding dimensions: Actively Open-minded Thinking (AOT) and Political Ideology (PI). AOT captures evidence-oriented belief revision, while PI captures identity-oriented interpretation of contested claims. Our central argument is that these traits shape resilience not only through their individual effects, but also through how they are distributed and coupled within a community. A population with many high-AOT agents may be more capable of questioning misinformation, but whether questioning leads to correction, rejection, or motivated resistance may depend on the ideological alignment of the claim. We therefore examine how different AOT--PI compositions shape collective recovery after misinformation shocks.

To study this process, we propose \textsc{CoSim}, a controlled LLM-agent simulation framework for examining how misinformation resilience emerges from psychologically heterogeneous communities. \textsc{CoSim} constructs synthetic communities with systematically varied AOT and PI profiles, exposes them to credible misinformation shocks, and models how agents interact, update trust, and transition across stances over time. The framework further supports multiple intervention strategies, including accuracy prompts, source warnings, fact checking, and persuasion, enabling controlled comparison of how interventions reshape aggregate recovery trajectories and the stance-transition pathways underlying them.

We organize the empirical analysis around three questions. First, we examine how AOT–PI population compositions shape two dimensions of resilience: robustness, or resistance to cumulative misinformation trust, and recovery, or the ability to correct misinformation-induced trust after its peak. Second, to explain why some AOT–PI groups are more resilient than others, we analyze the stance dynamics behind these trust trajectories: whether agents question the claim before trust peaks, whether they later reject or correct it, and whether they withdraw prior support after trust peaks. Third, we compare how accuracy prompts, persuasion, fact checking, and source warnings shift communities in the robustness–recovery space and alter stance transitions before and after the trust peak.

Our results show that misinformation resilience is best understood as a process of resistance and recovery rather than as a single measure of belief accuracy. Communities with higher AOT maintain lower cumulative trust in misinformation and are more likely to reduce trust after it peaks. PI shapes the recovery stage: more ideologically moderate communities recover more reliably, whereas polarized communities are more likely to retain residual support after exposure. The stance-level analysis shows that resilience depends on whether initial doubt becomes correction. Communities recover when agents move from questioning the claim’s authenticity to denying or correcting it and withdrawing prior support; they fail to recover when questioning remains unresolved or support persists. Interventions operate through the same pathway. Accuracy prompts mainly increase early caution, while persuasion and fact checking more directly support post-peak correction by reducing continued support and increasing explicit denial. Taken together, this work develops a community-level account of misinformation resilience, shows how the joint distribution of AOT and PI shapes resistance and recovery, and compares interventions by identifying the behavioral pathways through which collective correction succeeds or stalls.

Overall, this work makes three contributions:
\begin{itemize}
    \item We model misinformation resilience as a community-level dynamic process, emphasizing recovery dynamics rather than static individual belief outcomes.

    \item We characterize how population-level distributions and couplings of AOT and PI shape collective recovery through heterogeneous interaction patterns and stance transition dynamics.

    \item Building upon our developed technical framework with careful persona calibration and credible misinformation collections, we are able to provide a mechanism-level analysis of how psychological distributions and interventions reshape post-shock recovery trajectories by changing the conversion pathways among support, uncertainty, rejection, and correction.
\end{itemize}

\begin{figure*}[t]
    \centering
    \includegraphics[width=0.98\textwidth]{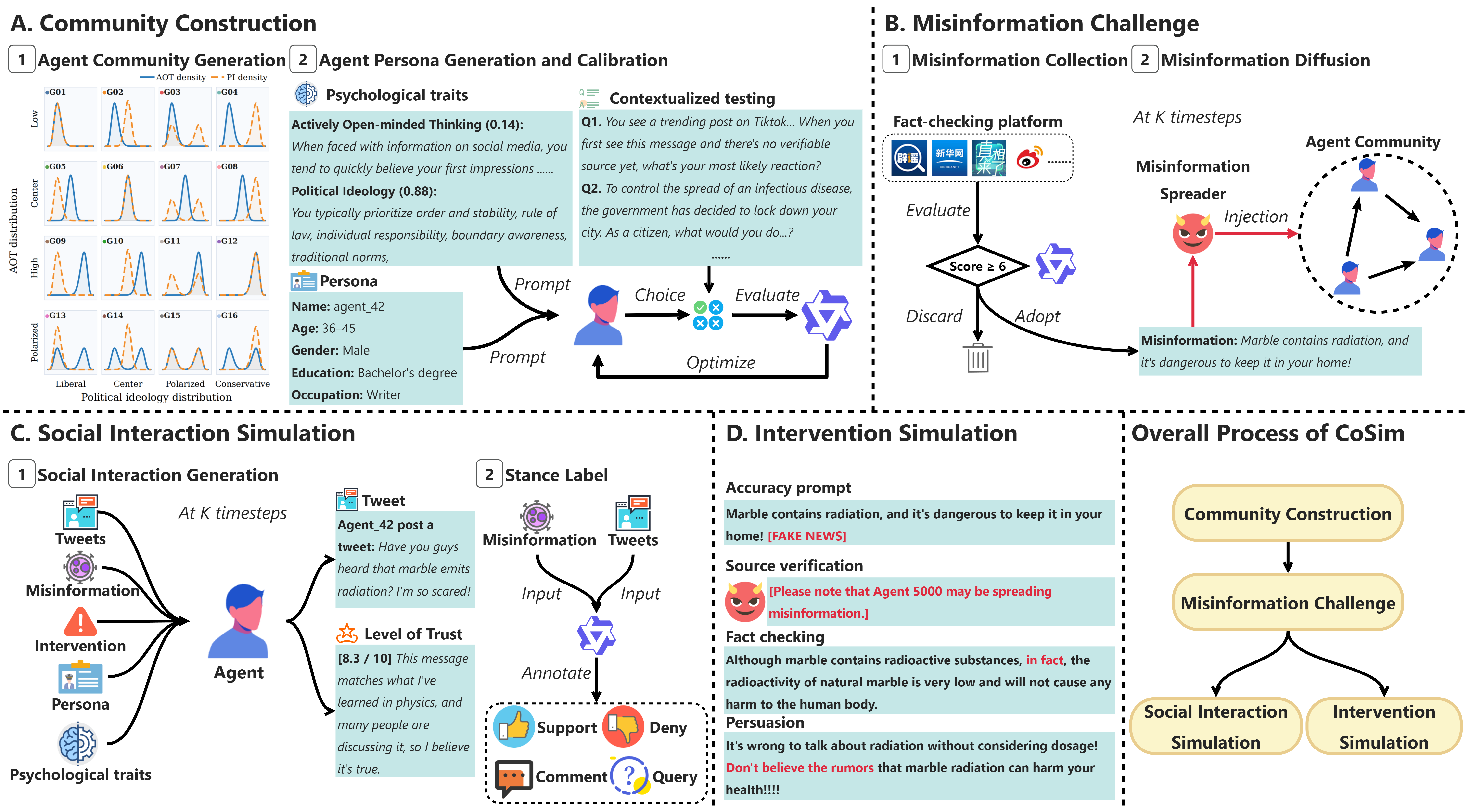}
    \caption{
        \textbf{Overview of CoSim.}
        The framework consists of four stages:
        \textbf{(A) community construction}, where psychological trait scores are sampled from predefined community distributions and mapped into calibrated persona prompts;
        \textbf{(B) misinformation challenge}, where verified misinformation cases are collected and injected into the community by a spreader agent;
        \textbf{(C) social interaction simulation}, where agents interact, update trust, and express stances over multiple timesteps; and
        \textbf{(D) intervention simulation}, where different mitigation strategies are introduced to evaluate their effects on misinformation resilience and recovery.
        }
    \label{fig:CoSim}
\end{figure*}
\section{Related Work}

\subsection{Misinformation Susceptibility and Intervention}

Misinformation susceptibility is shaped by both cognitive and social factors, including reasoning style, Actively Open-minded Thinking, prior belief, source credibility, Political Ideology, and attention to accuracy \citep{pennycook2019lazy,bago2020fake,pennycook2021psychology,ecker2022psychological,sultan2024susceptibility,biddlestone2025norm}. AOT is especially relevant to evidence processing because it reflects whether individuals are willing to consider belief-incongruent information and revise prior judgments, whereas PI and partisan identity shape whether a claim is perceived as credible, identity-consistent, or socially useful \citep{vanbavel2018partisan,osmundsen2021partisan,roozenbeek2022susceptibility}. Recent work further shows that misinformation resilience should not be reduced to blanket skepticism, since false-claim rejection may reflect either accurate truth discrimination or generalized distrust \citep{sultan2024susceptibility}. This distinction motivates our focus on trust dynamics, robustness, and recovery rather than simple disbelief.

Intervention studies have examined how misinformation response can be improved through prebunking, inoculation, accuracy prompts, warning labels, factual corrections, and user correction \citep{roozenbeek2019fake,pennycook2021shifting,roozenbeek2022psychological,bode2024user,altay2025small}. These studies show that misinformation susceptibility is malleable, but most focus on individual outcomes such as belief accuracy, perceived credibility, or sharing intention. Less is known about how resilience emerges in communities where heterogeneous users repeatedly observe, question, correct, and influence one another over time. This gap motivates a community-level approach that links psychological composition, interaction dynamics, and intervention effects.

\subsection{LLM Agent Simulation for Misinformation Resilience}

LLM agents provide a controlled way to simulate social interaction because they can express attitudes, maintain memory, generate natural language actions, and interact in structured environments \citep{park2023generative,park2024generative,mou2024individual}. In misinformation research, earlier computational work mainly modeled diffusion and exposure patterns, showing how false information spreads through networks and how virality, bots, and low-credibility content shape information ecosystems \citep{vosoughi2018spread,shao2018spread,allen2020evaluating}. Recent LLM agent studies extend this line by modeling fake news diffusion, moderation, content transformation, demographic-aware persuasion, and persona-conditioned propagation \citep{li2024newsdiffusion,liu2025mosaic,liu2025stepwise,borah2026persuasion,brian2025mpcg,maurya2025simulating,composta2025simulating}. Other studies evaluate whether LLM agents can approximate human misinformation susceptibility, while also warning that such simulations may overstate attitude effects or underweight personal network factors \citep{pratelli2025evaluating,choi2026overstating}. Most closely related to our psychological composition setting, \citet{borah2026belief} shows that belief priors improve LLM-based simulation of demographic misinformation susceptibility, suggesting that latent belief structures matter beyond demographic labels.

CoSim, our proposed LLM-agent framework for studying misinformation resilience, builds on this literature but shifts the unit of analysis from individual susceptibility, message transformation, and diffusion outcomes to community resilience. Rather than asking only whether agents believe, spread, or are persuaded by misinformation, CoSim examines how psychologically composed communities withstand credible misinformation shocks and recover through repeated interaction. By constructing communities from distributions of AOT and PI, CoSim links psychological composition to robustness, recovery, and behavioral mechanisms such as questioning, denial, support withdrawal, and social correction.

\section{The CoSim Framework}
\label{The CoSim Framework}

CoSim models community resilience to misinformation as a dynamic process rather than a static belief judgment. Building on recent social simulation studies of misinformation diffusion and conversational dynamics~\citep{liu2025stepwise,jain2025modeling}, CoSim examines how agent communities absorb credible misinformation shocks, interact under repeated exposure, update trust and stance, and respond to external interventions. The framework is implemented on top of AgentSociety~\citep{piao2025agentsociety}, which provides the underlying multi-agent simulation infrastructure for agent scheduling, memory management, and interaction orchestration.

As shown in Figure~\ref{fig:CoSim}, CoSim consists of four modules: Community Construction, Misinformation Challenge, Social Interaction Simulation, and Intervention Simulation. It first constructs psychologically controlled agent communities and injects credible false claims as misinformation challenges. Agents then observe misinformation and peer responses, generate trust judgments and behavioral reactions, and update their memories over multiple rounds. Finally, interventions are introduced to evaluate whether they reduce misinformation uptake and support post-exposure recovery.


\begin{figure*}[t]
    \centering
    \includegraphics[width=0.95\linewidth]{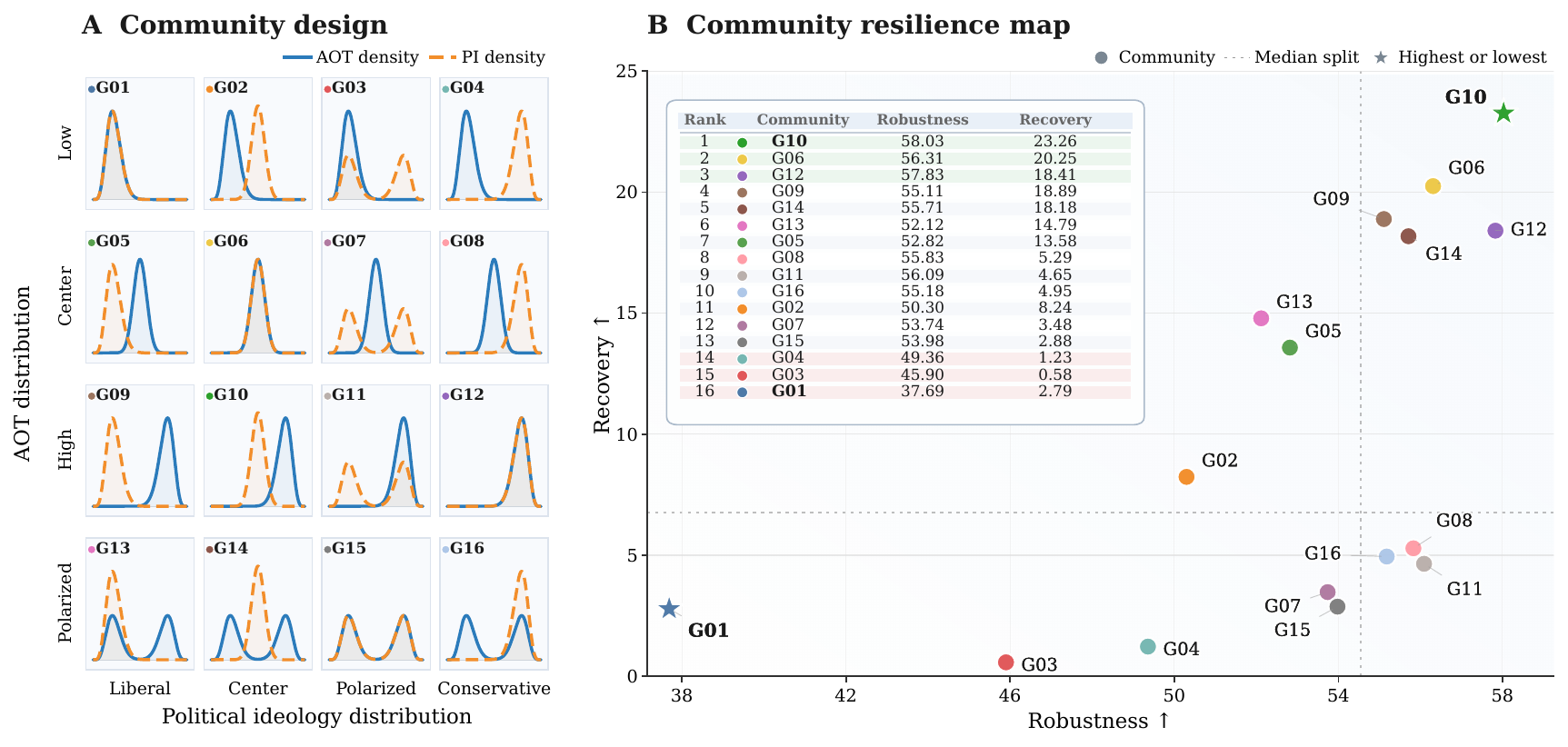}
    \caption{
        \textbf{Community design and resilience outcomes.}
        \textbf{(A) Community design} crosses four AOT distributions with four PI distributions; blue solid and orange dashed curves show the corresponding AOT and PI densities.
        \textbf{(B) Resilience map} plots robustness and recovery scores, where higher values indicate stronger resilience.
        Dashed lines mark median splits, stars mark the highest and lowest composite scores, and the inset ranks communities by the mean of min--max-normalized robustness and recovery.
    }
    \label{fig:rq1}
\end{figure*}

\begin{figure*}[htbp]
    \centering
    \includegraphics[width=0.95\linewidth]{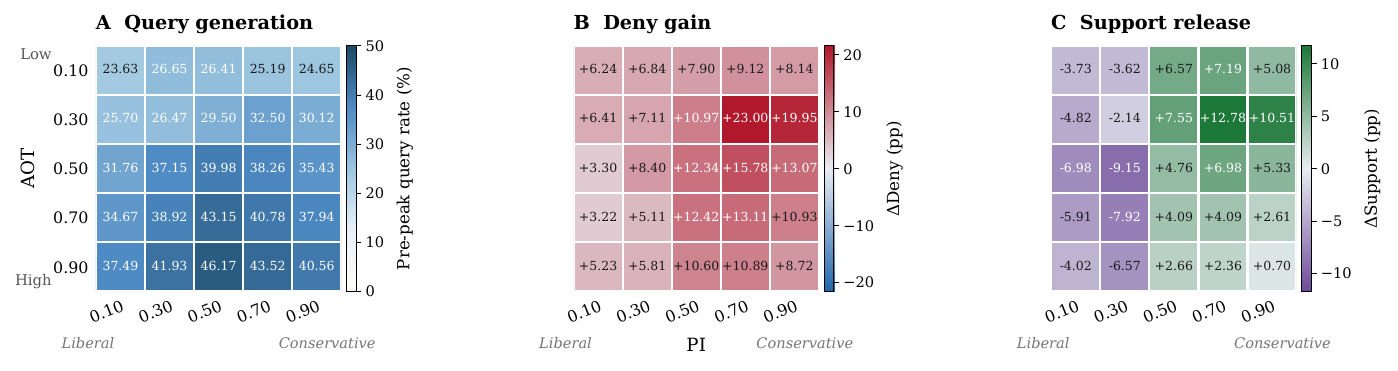}
    \caption{
        \textbf{Cell-level behavioral mechanisms across AOT-by-PI profiles.}
        \textbf{(A) Query generation} measures the pre-peak tendency to question misinformation-related information.
        \textbf{(B) Deny gain} measures the post-peak conversion from scrutiny to explicit rejection.
        \textbf{(C) Support release} measures the post-peak reduction in misinformation support, with positive values indicating support withdrawal and negative values indicating persistent or increased support.
        Read \textbf{vertically}, the maps reveal the AOT effect on query generation; read \textbf{horizontally}, they reveal the PI effect on query-to-deny conversion and support release.
    }
    \label{fig:rq2_mechanism}
\end{figure*}
\subsection{Community Construction}
\label{sec:community_construction}

CoSim separates community construction into two linked steps: sampling numerical psychological traits to define community composition, and converting these trait scores into calibrated persona prompts that instantiate LLM agents.

\paragraph{Agent community generation.}
CoSim first constructs synthetic communities by sampling numerical psychological traits for each agent. Each community is indexed by an AOT profile \(a\) and a PI profile \(p\). For agent \(u_i\), we sample
\begin{equation}
\alpha_i\sim D^{AOT}_{a}, \qquad
\pi_i\sim D^{PI}_{p},
\end{equation}
where \(\alpha_i \in [0,1]\) denotes the agent's AOT score and \(\pi_i \in [0,1]\) denotes the agent's PI score. We encode PI such that \(\pi_i=0\) denotes the liberal pole and \(\pi_i=1\) denotes the conservative pole.

Given the sampled trait scores, each agent is instantiated through a persona prompt:
\begin{equation}
P_i=\mathcal{T}_{\theta}(\alpha_i,\pi_i,b_i),
\end{equation}
where \(\mathcal{T}_{\theta}\) is a parameterized persona template and \(b_i\) denotes controlled background attributes. The resulting community is defined as
\begin{equation}
\mathcal{C}^{(a,p)}=\{u_i=(P_i,\alpha_i,\pi_i)\}_{i=1}^{N}.
\end{equation}

For each trait, CoSim defines four distributional profiles. AOT includes low, center, high, and polarized profiles, while PI includes liberal, center, conservative, and polarized profiles. Their Cartesian product yields 16 community types. This design does not aim to reproduce a specific empirical population. Instead, it provides controlled synthetic communities whose trait distributions can be systematically varied while the simulation protocol remains fixed.

\paragraph{Persona prompt generation and calibration.}
To reduce prompt-induced trait drift, CoSim calibrates the persona template using a diagnostic questionnaire set \(\mathcal{Q}\). For each agent, the LLM generates responses to the diagnostic questionnaire conditioned on the persona prompt:
\begin{equation}
\mathbf{y}_i^{(s)}
=
f_{\phi}(\mathcal{Q}\mid P_i^{(s)}),
\end{equation}
where \(s\) denotes the calibration iteration, \(f_{\phi}\) denotes the LLM backbone, and \(\mathbf{y}_i^{(s)}\) denotes the response vector. A rule based mapper \(S(\cdot)\) then estimates the realized traits:
\begin{equation}
(\hat{\alpha}_i^{(s)},\hat{\pi}_i^{(s)})=S(\mathbf{y}_i^{(s)}).
\end{equation}

At each calibration iteration, we measure the discrepancy between target and realized traits:
\begin{equation}
\mathcal{L}(\theta^{(s)})=
\frac{1}{N}\sum_{i=1}^{N}
\left(
|\alpha_i-\hat{\alpha}_i^{(s)}|
+
|\pi_i-\hat{\pi}_i^{(s)}|
\right).
\end{equation}
An LLM-based optimizer revises the persona template according to diagnostic errors. We run this calibration procedure for \(R=10\) iterations and select the template with the lowest diagnostic loss:
\begin{equation}
\theta^{*}
=
\arg\min_{\theta^{(s)}:s\in\{1,\ldots,R\}}
\mathcal{L}(\theta^{(s)}),
\qquad R=10.
\end{equation}
The calibrated template \(\mathcal{T}_{\theta^{*}}\) is then fixed and used for all subsequent simulations. Details of the diagnostic questionnaire, scoring rules, and calibration procedure are provided in Appendix.


\subsection{Misinformation Challenge}
\label{sec:misinformation_challenge}

The misinformation challenge is designed to provide credible but externally refuted claims. CoSim builds a pool of misinformation from 5,194 misinformation records published over the past seven years and collected from four authoritative Chinese fact checking and public communication sources. Four LLM backbones score each candidate claim for perceived credibility, and 105 externally refuted claims with an average credibility score above 6 are retained. The final pool covers multiple domains, years, credibility intervals, and claim lengths, allowing us to test whether different communities can resist and recover from comparable misinformation shocks. Detailed source descriptions, filtering criteria, and pool distributions are provided in the appendix.

Each selected misinformation claim \(r\) contains claim content, perceived credibility, and source identity.  All communities are exposed to the same pool of credible false claims. After a claim is selected, misinformation is injected into the community through a spreading agent. At each round \(t\), CoSim independently samples a directly exposed subset \(\mathcal{E}_t\subseteq \mathcal{C}^{(a,p)}\). We set the exposure ratio to \(\rho=0.1\), so that approximately 10\% of agents receive the misinformation claim directly in each round:
\begin{equation}
\lvert \mathcal{E}_t \rvert = \lceil \rho N \rceil, \quad \rho=0.1.
\end{equation}
The exposed set is sampled uniformly at random from all subsets of \(\mathcal{C}^{(a,p)}\) with size \(\lceil \rho N \rceil\):
\begin{equation}
\mathcal{E}_t
\sim
\mathrm{Unif}
\left(
\left\{
\mathcal{E}\subseteq \mathcal{C}^{(a,p)}
:
\lvert \mathcal{E} \rvert = \lceil \rho N \rceil
\right\}
\right).
\end{equation}
Agents in \(\mathcal{E}_t\) receive the misinformation claim directly, while other agents may encounter it indirectly through local peer messages. The direct exposure input is defined as
\begin{equation}
x_{i,t}=
\begin{cases}
r, & u_i\in \mathcal{E}_t,\\
\varnothing, & u_i\notin \mathcal{E}_t.
\end{cases}
\end{equation}

\begin{figure*}[t]
    \centering
    \includegraphics[width=0.95\linewidth]{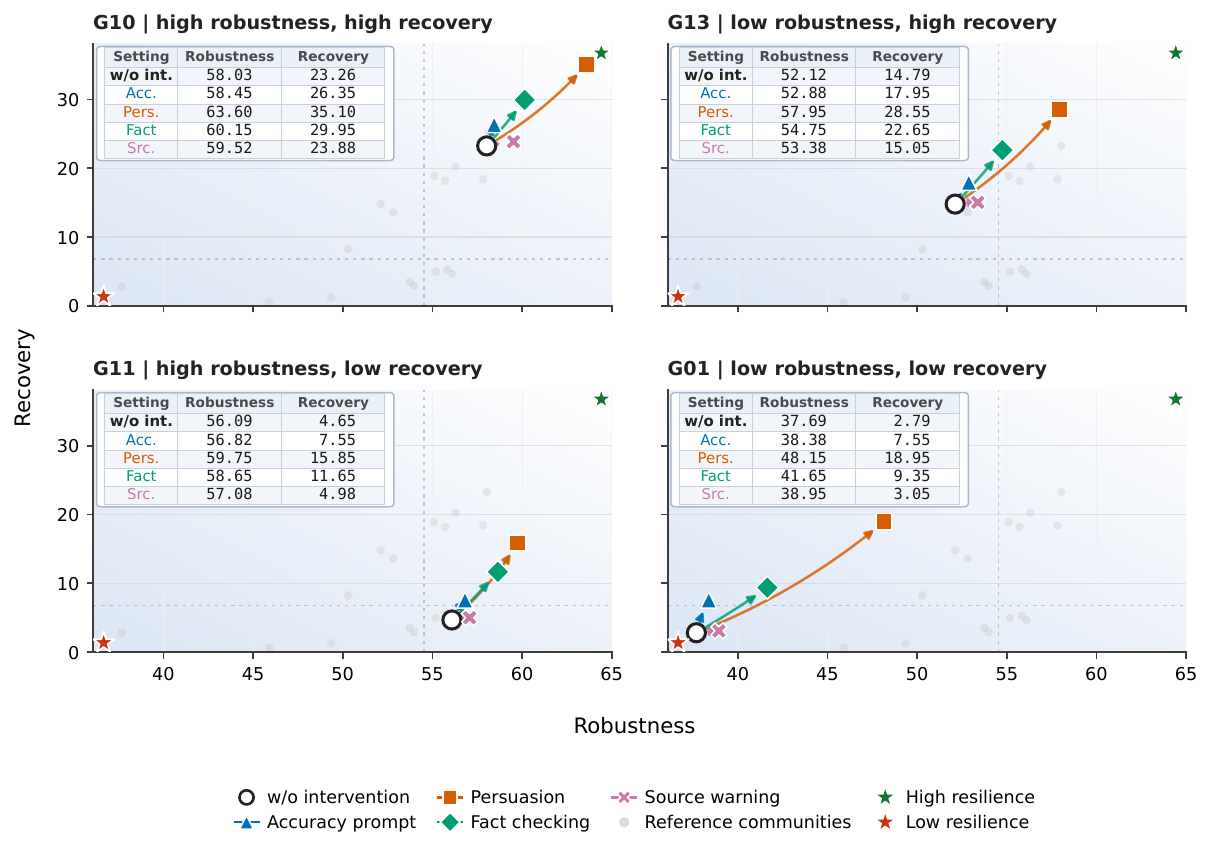}
    \caption{
    \textbf{Intervention effects in the robustness--recovery space.}
    We evaluate four representative communities selected from RQ1: G10, G13, G11, and G01, which cover the four combinations of high/low robustness and high/low recovery.
    Each panel compares the control condition with four interventions: accuracy prompt, persuasion, fact checking, and source warning.
    The x-axis reports robustness, where higher values indicate lower cumulative misinformation trust; the y-axis reports recovery, where higher values indicate stronger post-shock correction.
    Gray points show the 16 reference communities from RQ1, stars mark the high- and low-resilience regions, and arrows indicate intervention-induced movement from the control condition.
    Effective interventions move communities toward the upper-right region.
    }
    \label{fig:rq3_outcome_space}
\end{figure*}

\begin{figure}[t]
    \centering
    \includegraphics[width=0.95\linewidth]{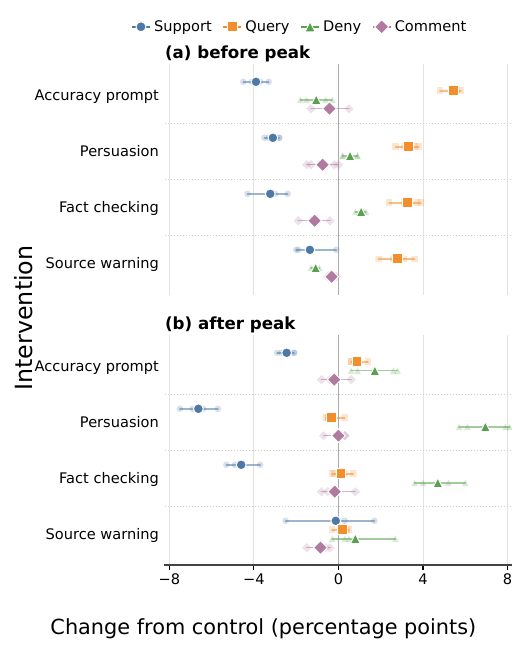}
    \caption{
    \textbf{Intervention-induced stance shifts before and after the trust peak.}
    Values denote percentage-point changes relative to the control condition, averaged across the representative communities in Figure~\ref{fig:rq3_outcome_space}.
    Panel (a) captures the pre-peak uptake stage, while Panel (b) captures the post-peak adjustment stage.
    Negative support values indicate reduced misinformation endorsement, whereas positive query and denial values indicate stronger authenticity questioning and explicit rejection.
    Accuracy prompts mainly induce early caution; persuasion and fact checking more effectively convert post-peak support into denial.
    }
    \label{fig:rq3_stance_shift}
\end{figure}


\subsection{Social Interaction Simulation}
\label{sec:social_interaction}

CoSim models misinformation response as a repeated social process over a social network \(\mathcal{G}\). For each agent \(u_i\), we define its observable peer set as
\begin{equation}
\mathcal{N}_i=\{u_j:(u_j,u_i)\in\mathcal{G}\}.
\end{equation}
At round \(t\), the local peer context is constructed from messages generated by observable peers in the previous round:
\begin{equation}
\mathcal{M}_{i,t}=\{m_{j,t-1}:u_j\in\mathcal{N}_i\}.
\end{equation}

At each round, agent \(u_i\) observes its persona, prior memory, direct exposure input, local peer context, and an optional intervention signal. The agent then generates a trust judgment, an opinion expression, and a public message:
\begin{equation}
(\tau_{i,t}, o_{i,t}, m_{i,t})
=
f_{\phi}(P_i,H_{i,t-1},x_{i,t},\mathcal{M}_{i,t},z_{i,t}).
\end{equation}
Here, \(\tau_{i,t}\in[0,1]\) denotes trust in the misinformation, \(o_{i,t}\) denotes the agent's private opinion expression, \(m_{i,t}\) denotes the generated public message, \(x_{i,t}\) denotes direct exposure to the misinformation, \(\mathcal{M}_{i,t}\) denotes local peer context, and \(z_{i,t}\) denotes the intervention signal. When no intervention is applied, \(z_{i,t}=\varnothing\).

The generated message is further annotated into one of four behavioral stances:
\begin{equation}
s_{i,t}=A(m_{i,t}), \qquad
s_{i,t}\in\{\text{support},\text{deny},\text{query},\text{comment}\},
\end{equation}
where \(A(\cdot)\) denotes the stance annotator. This label space follows misinformation stance detection research~\citep{derczynski2017semeval,gorrell2019rumoureval}. Support indicates acceptance or amplification of the misinformation, deny indicates rejection or correction, query captures uncertainty and verification demand, and comment denotes peripheral or noncommittal discussion.

After response generation and annotation, the agent memory is updated by appending the current interaction record:
\begin{equation}
H_{i,t}
=
H_{i,t-1}
\oplus
\left(
x_{i,t},
\mathcal{M}_{i,t},
\tau_{i,t},
o_{i,t},
s_{i,t},
m_{i,t}
\right).
\end{equation}
Here, \(\oplus\) denotes chronological appending, preserving the temporal order of exposures, peer contexts, trust judgments, opinion expressions, stance labels, and generated messages.

The community-level average trust trajectory for community \(g\) under model backbone \(m\) is computed as
\begin{equation}
\bar{\tau}_{g,m}(t)
=
\frac{1}{N}
\sum_{i=1}^{N}
\tau_{i,t}^{(g,m)}.
\end{equation}
This trajectory records how the community moves from initial misinformation exposure toward belief persistence, recovery, or generalized distrust.


\subsection{Intervention Simulation}
\label{sec:intervention_simulation}

CoSim models interventions as modifications to the information context rather than changes to agent traits or decision rules. At round \(t\), an intervention operator produces an agent specific signal:
\begin{equation}
z_{i,t}=\mathcal{I}_k(r,u_i,t,H_{i,t-1}),
\end{equation}
where \(k\) denotes the intervention type. When no intervention is applied, \(z_{i,t}=\varnothing\). This design allows different interventions to enter the response process as contextual signals while keeping the agent persona and the LLM decision rule fixed.

\paragraph{Accuracy prompt.}
A short warning cue is appended to each misinformation message, indicating that the claim may be false. This intervention targets the earliest judgment stage by making accuracy salient before the agent forms a trust judgment or public reaction. It follows prior evidence that accuracy prompts can reduce misinformation sharing by redirecting attention toward the truthfulness of online content~\citep{pennycook2022accuracy}.

\paragraph{Persuasion.}
Directly exposed agents receive a corrective persuasive message that explains why the claim may be misleading and encourages reconsideration. Compared with the accuracy prompt, persuasion provides a richer corrective frame after exposure. This design is motivated by correction research showing that misinformation can continue to influence belief even after correction, and that effective correction often requires more substantive explanatory content rather than a simple warning cue~\citep{walter2020meta}.

\paragraph{Fact checking.}
Fact checking is modeled as an active verification process rather than automatic evidence delivery. In this condition, the intervention first informs the agent that factual evidence can be requested. The agent then decides whether to verify the claim:
\begin{equation}
q_{i,t}
=
g_{\phi}(P_i,H_{i,t-1},x_{i,t},\mathcal{M}_{i,t},a_r),
\end{equation}
where \(q_{i,t}\in\{0,1\}\) denotes whether agent \(u_i\) requests factual evidence, and \(a_r\) denotes the availability of fact checking for claim \(r\). The intervention signal is then defined as
\begin{equation}
z_{i,t}^{FC}
=
\begin{cases}
e_r, & q_{i,t}=1,\\
a_r, & q_{i,t}=0,
\end{cases}
\end{equation}
where \(e_r\) denotes factual evidence about the claim's refuted status. Thus, corrective information is delivered only to agents who actively request verification. This intervention tests whether recovery depends on active verification rather than passive correction. It is inspired by community-based fact checking systems, where corrective context becomes effective when users engage with evidence and use it to reassess misleading content~\citep{chuai2026community}.

\paragraph{Source warning.}
The true identity of the misinformation spreading agent is revealed. This intervention tests whether source transparency helps agents recalibrate trust and reduce continued belief in the false claim. This design is grounded in evidence that source credibility information can improve truth discernment and reduce engagement with misinformation online~\citep{prike2024source}.

Together, these interventions probe four mechanisms of community resilience: accuracy salience, corrective persuasion, active verification, and source distrust.

\section{Evaluation}
\label{section:evaluation}

We organize evaluation around three questions: whether psychological composition shapes resilience, which stance dynamics explain resilience, and which interventions improve recovery.

\textbf{RQ1: How do psychological distributions shape community responses to credible misinformation shocks?}

\textbf{RQ2: What behavioral processes distinguish resilient communities from communities that remain trapped in false belief or generalized distrust?}

\textbf{RQ3: Which intervention strategies most effectively improve community recovery under misinformation exposure?}

\subsection{Experimental Settings}

We construct 16 community types by crossing four AOT distributions with four PI distributions. Each community contains \(N=200\) agents connected by a fixed small-world network with average degree \(\bar{d}=6\) and rewiring probability \(p_{\mathrm{rewire}}=0.1\). Each simulation runs for \(T=10\) time steps. At each round, CoSim randomly exposes 10\% of agents to the misinformation claim, while other agents may encounter related content through messages from their observable peers. CoSim records agent-level trust judgments, opinion expressions, stance labels, and generated messages across rounds. Unless otherwise specified, all LLMs use the same decoding settings, with temperature \(=0.5\). The main analysis reports Qwen3-4B~\cite{yang2025qwen3} results, with Qwen2.5-3B~\cite{bai2023qwen}, Phi-4-mini~\cite{abouelenin2025phi}, and Gemma3-4B~\cite{gemma2025technical} used for backbone robustness checks.

\subsection{RQ1}
RQ1 evaluates whether different AOT and PI distributions produce distinguishable community responses under credible misinformation shocks. For each community, we summarize the misinformation trust trajectory using two indicators: robustness and recovery. Both metrics are reported on a 0 to 100 scale, where higher values indicate stronger resilience.

Let \(\bar{\tau}_{g,m}(t)\) denote the average misinformation trust of community \(g\) under model \(m\) at simulation time \(t\). Since higher trust indicates stronger belief in misinformation, we first compute normalized misinformation exposure as the area under the trust curve:
\begin{equation}
\mathrm{Exposure}_{g,m}
=
100 \times
\frac{1}{T \tau_{\max}}
\int_{0}^{T}
\bar{\tau}_{g,m}(t)\,dt .
\end{equation}

In discrete simulations, this exposure is approximated using the trapezoidal rule. Robustness is then defined as the complement of exposure:
\begin{equation}
\mathrm{Robustness}_{g,m}
=
100
-
\mathrm{Exposure}_{g,m}.
\end{equation}
A higher robustness score, therefore, indicates lower cumulative belief in misinformation.

Recovery captures how much of the misinformation-induced trust increase is corrected by the end of the simulation. We define
\begin{equation}
\tau^{peak}_{g,m}=\max_t \bar{\tau}_{g,m,t}, \quad
\tau^{0}_{g,m}=\bar{\tau}_{g,m,0}, \quad
\tau^{T}_{g,m}=\bar{\tau}_{g,m,T}.
\end{equation}
Recovery is then computed as
\begin{equation}
\mathrm{Recovery}_{g,m}
=
100 \times
\frac{
\tau^{peak}_{g,m}
-
\tau^{T}_{g,m}
}{
\tau^{peak}_{g,m}
-
\tau^{0}_{g,m}
+\epsilon
}.
\end{equation}
Here, \(\epsilon\) is a small constant used to avoid division by zero. A higher recovery score indicates stronger post-shock belief correction.

\subsubsection{Finding RQ1.}

Figure~\ref{fig:rq1} shows that resilience varies systematically across community designs. Under Qwen3-4B, higher AOT generally improves robustness: Low AOT communities remain the least robust, with G01 reaching only 37.69, whereas High AOT communities achieve consistently higher robustness, including G10 at 58.03 and G12 at 57.83. PI further differentiates these outcomes. PI further differentiates these outcomes. Center PI provides the most balanced resilience profile, whereas Conservative PI is associated with stronger resistance in several settings and Polarized PI consistently weakens recovery.

Recovery exhibits a more polarized pattern. Center PI yields the strongest correction capacity within each AOT block, with G06, G10, and G14 reaching 20.25, 23.26, and 18.18, respectively. In contrast, Polarized PI consistently weakens recovery, as shown by G03, G07, G11, and G15, whose recovery values remain below 5.0. This suggests that ideological polarization is not equivalent to ideological balance: it may limit exposure in some cases, but it suppresses collective correction after the shock.

Taken together, G10, combining High AOT with Center PI, achieves the best overall profile, ranking first in both robustness and recovery. These results indicate that community resilience is better understood as a joint outcome of resistance and correction: higher AOT improves both dimensions, Center PI provides the most stable balance, Conservative PI mainly supports resistance, and Polarized PI constrains recovery. Additional results across model backbones are reported in the Appendix and show the same qualitative pattern.

\subsection{RQ2}

RQ2 examines how community resilience emerges from stance dynamics rather than aggregate trust curves alone. We partition agents into AOT by PI cells and use the trust peak to separate misinformation uptake from post-peak adjustment. For each cell, we compute query generation, deny gain, and support release, capturing questioning, rejection conversion, and support withdrawal.

\subsubsection{Finding RQ2.}

Figure~\ref{fig:rq2_mechanism} should be read in two directions. Reading Panel A vertically reveals the AOT effect: higher AOT consistently increases the supply of pre-peak scrutiny. The average query rate rises from 25.31\% in the lowest AOT row to 41.93\% in the highest AOT row, with the strongest cell reaching 46.17\%. This helps explain why low AOT communities such as G01 and G03 stay near the bottom of Figure~\ref{fig:rq1}, with robustness and recovery scores of 37.69/2.79 and 45.90/0.58, respectively.

Reading Panels B and C horizontally reveals the PI effect. Conservative-leaning cells show stronger query to denial conversion and stronger support release. Deny gain increases from 4.88 percentage points in the most liberal PI column to 14.38 percentage points in the conservative-leaning column, while support release shifts from negative values in liberal cells to positive values in center and conservative cells. Thus, PI determines whether scrutiny remains unresolved or becomes a corrective stance change.

This mechanism accounts for the high resilience region in Figure~\ref{fig:rq1}. G10, G06, and G12 combine sufficient query generation with effective post-peak conversion, producing strong robustness and recovery scores of 58.03/23.26, 56.31/20.25, and 57.83/18.41. In contrast, G08, G11, and G16 show moderate to high robustness but weak recovery, suggesting that questioning information authenticity may suppress cumulative misinformation trust, whereas full recovery requires scrutiny to be converted into denial and support release.

\subsection{RQ3}

RQ3 evaluates whether external interventions can strengthen community resilience after a credible misinformation shock. Based on the outcome space identified in RQ1, we select four representative communities: G10, G13, G11, and G01, covering the four combinations of high or low robustness and high or low recovery. For each community, we compare the control condition with accuracy prompt, persuasion, fact checking, and source warning. Figure~\ref{fig:rq3_outcome_space} reports outcome shifts in the robustness and recovery plane, while Figure~\ref{fig:rq3_stance_shift} decomposes these shifts into stance changes before and after the trust peak.

\subsubsection{Finding RQ3.}

Figure~\ref{fig:rq3_outcome_space} shows that persuasion is the strongest and most stable intervention across representative community types. Because higher robustness indicates lower cumulative misinformation trust and higher recovery indicates stronger post-shock correction, effective interventions should move communities toward the upper right region. Persuasion produces the largest joint improvement in all four representative communities. The effect is particularly salient for the weakest community, G01, where robustness increases from 37.69 to 48.15 and recovery from 2.79 to 18.95. Similar gains appear in G13, where persuasion improves robustness from 52.12 to 57.95 and recovery from 14.79 to 28.55. Persuasion also substantially improves G10 and G11, raising G10 from 58.03 to 63.60 in robustness and from 23.26 to 35.10 in recovery, and raising G11 from 56.09 to 59.75 in robustness and from 4.65 to 15.85 in recovery. By contrast, accuracy prompts and source warnings produce smaller shifts, with source warnings remaining close to the control condition in most panels. Fact checking also improves outcomes, but its gains are generally weaker than persuasion. These results suggest that intervention effectiveness depends on the richness and specificity of corrective information: source warnings provide a credibility cue, accuracy prompts offer a general caution signal, fact checking supplies corrective evidence, and persuasion further builds on evidence to make correction easier to accept.

Figure~\ref{fig:rq3_stance_shift} explains the behavioral mechanism behind these outcome differences. Accuracy prompt produces the strongest early attention effect, reducing pre-peak support by 3.90 percentage points and increasing pre-peak query by 5.43 percentage points. However, this early hesitation translates less strongly into post-peak correction: after the peak, the accuracy prompt reduces support by 2.45 percentage points and increases denial by 1.73 percentage points. In comparison, persuasion and fact checking show stronger post-peak conversion. Persuasion reduces post-peak support by 6.62 percentage points and increases denial by 6.95 percentage points, while fact checking reduces support by 4.60 percentage points and increases denial by 4.70 percentage points. Thus, the most effective interventions are not those that merely make agents uncertain, but those that convert uncertainty into explicit rejection of the false claim.

\section{Conclusion}
This paper examined why some communities recover from misinformation while others retain support for false claims. Using controlled LLM-agent simulations, we find that higher AOT improves both resistance to misinformation uptake and recovery after trust peaks. PI shapes the recovery pathway: ideologically moderate communities recover more reliably, while polarized communities are more likely to retain residual support. At the stance level, recovery depends on whether agents move from questioning a claim to denying or correcting it and withdrawing prior support. Intervention results further show that persuasion and fact checking are more effective at supporting post-peak correction, whereas accuracy prompts mainly induce early caution and source warnings have weaker effects.

The study also has limitations. First, the PI results should be interpreted cautiously. Conservative-leaning PI sometimes improves robustness, while liberal-leaning PI can improve recovery, but these patterns are based on a Chinese misinformation pool and an expert-coded liberal–conservative mapping. They should not be generalized without further studies across different political, cultural, and media contexts. Second, LLM agents are not real human users. Although they can maintain personas, memory, and multi-round interactions, their behavior remains synthetic. Future work should compare these trust and stance dynamics with human-subject experiments or platform data. Third, the current robustness tests are limited. Misinformation resilience may also depend on network structure, network position, and exposure patterns. Future studies should test more diverse misinformation settings, including repeated and personalized exposure, different network structures, and more realistic intervention timing.

\bibliography{aaai2026}

\section{Ethics Checklist}

\begin{enumerate}

\item For most authors...
\begin{enumerate}
    \item Would answering this research question advance science without violating social contracts, such as violating privacy norms, perpetuating unfair profiling, exacerbating the socio-economic divide, or implying disrespect to societies or cultures?
    \answerYes{The study uses controlled LLM agent simulations to analyze misinformation resilience. It does not profile real individuals or intervene in real online communities.}

    \item Do your main claims in the abstract and introduction accurately reflect the paper's contributions and scope?
    \answerYes{The abstract and introduction describe the paper's main scope: community-level misinformation resilience, psychological composition, stance dynamics, and intervention effects.}

    \item Do you clarify how the proposed methodological approach is appropriate for the claims made?
    \answerYes{We explain the motivation for controlled LLM agent simulation in the Introduction, CoSim Framework, and Experimental Settings.}

    \item Do you clarify what are possible artifacts in the data used, given population specific distributions?
    \answerYes{We discuss possible artifacts from synthetic communities, persona construction, selected misinformation stimuli, and LLM backbone behavior in Limitations and Ethical Considerations.}

    \item Did you describe the limitations of your work?
    \answerYes{Limitations are discussed in Limitations and Ethical Considerations.}

    \item Did you discuss any potential negative societal impacts of your work?
    \answerYes{Potential negative societal impacts are discussed in Limitations and Ethical Considerations.}

    \item Did you discuss any potential misuse of your work?
    \answerYes{We discuss possible misuse of misinformation simulations and intervention analysis in Limitations and Ethical Considerations.}

    \item Did you describe steps taken to prevent or mitigate potential negative outcomes of the research, such as data and model documentation, data anonymization, responsible release, access control, and the reproducibility of findings?
    \answerYes{We conduct the study in a closed simulation environment, report model versions and experimental settings, and provide prompts and configurations for reproducibility. Misinformation related materials are handled with responsible release considerations.}

    \item Have you read the ethics review guidelines and ensured that your paper conforms to them?
    \answerYes{Yes.}
\end{enumerate}

\item Additionally, if your study involves hypotheses testing...
\begin{enumerate}
    \item Did you clearly state the assumptions underlying all theoretical results?
    \answerNA{The paper does not present formal theoretical results or proofs. It is organized around research questions and controlled simulation experiments.}

    \item Have you provided justifications for all theoretical results?
    \answerNA{The paper does not present formal theoretical results or proofs.}

    \item Did you discuss competing hypotheses or theories that might challenge or complement your theoretical results?
    \answerYes{Although the paper is organized around research questions rather than formal hypotheses, we discuss alternative explanations related to LLM agent artifacts, persona construction, misinformation stimulus selection, ideological coding, network structure, and intervention idealization.}

    \item Have you considered alternative mechanisms or explanations that might account for the same outcomes observed in your study?
    \answerYes{We discuss alternative mechanisms including network position, exposure patterns, repeated exposure, source credibility, institutional trust, and intervention timing. These factors are discussed as limitations and future work.}

    \item Did you address potential biases or limitations in your theoretical framework?
    \answerYes{We discuss limitations of the AOT and PI construct design, possible construct transfer errors in the liberal and conservative coding scheme, and the limited external validation of LLM agent behavior.}

    \item Have you related your theoretical results to the existing literature in social science?
    \answerYes{The paper relates the simulation design and findings to prior work on misinformation susceptibility, Actively Open-minded Thinking, Political Ideology, fact checking, accuracy prompts, source warnings, correction, and community-based misinformation response.}

    \item Did you discuss the implications of your theoretical results for policy, practice, or further research in the social science domain?
    \answerYes{We discuss implications for studying community misinformation resilience as a dynamic process, while emphasizing that CoSim is a diagnostic framework rather than a direct policy prescription. Future work should validate simulated mechanisms against human behavior and platform data.}
\end{enumerate}

\item Additionally, if you are including theoretical proofs...
\begin{enumerate}
    \item Did you state the full set of assumptions of all theoretical results?
    \answerNA{The paper does not include theoretical proofs.}

    \item Did you include complete proofs of all theoretical results?
    \answerNA{The paper does not include theoretical proofs.}
\end{enumerate}

\item Additionally, if you ran machine learning experiments...
\begin{enumerate}

    \item Did you include the code, data, and instructions needed to reproduce the main experimental results, either in the supplemental material or as a URL?
    \answerNo{We do not include the full code at submission because the CoSim implementation is still under active development, including ongoing extensions for studying the phase transition properties of community misinformation resilience. To support transparency, the paper and appendix provide prompts, simulation settings, community construction procedures, intervention settings, evaluation metrics, and aggregate outputs needed to understand and inspect the main experimental results. After publication, implementation details can be requested by contacting the authors by email.}

    \item Did you specify all the training details, such as data splits, hyperparameters, and how they were chosen?
    \answerNA{We do not train or fine-tune models. We use existing LLM backbones for inference-based simulation.}

    \item Did you report error bars, for example, with respect to the random seed after running experiments multiple times?
    \answerYes{We run each key simulation experiment with five random seeds.}

    \item Did you include the total amount of compute and the type of resources used, such as type of GPUs, internal cluster, or cloud provider?
    \answerYes{We report the evaluated LLM backbones and the local inference environment in Experimental Settings and the appendix.}

    \item Do you justify how the proposed evaluation is sufficient and appropriate to the claims made?
    \answerYes{The evaluation is organized around three research questions covering community composition, behavioral mechanisms, and intervention effects.}

    \item Do you discuss what is ``the cost'' of misclassification and fault intolerance?
    \answerYes{The study uses LLM-based credibility scoring and stance annotation for simulation analysis rather than real-world decision making. Errors in these components may affect estimated mechanisms and aggregate resilience scores, but they do not trigger automated decisions about real users. We discuss these limitations in the appendix.}
\end{enumerate}

\item Additionally, if you are using existing assets, such as code, data, models, or curating and releasing new assets, \textbf{without compromising anonymity}...
\begin{enumerate}
    \item If your work uses existing assets, did you cite the creators?
    \answerYes{We cite the existing LLM backbones, simulation related work, and misinformation research used in the study.}

    \item Did you mention the license of the assets?
    \answerYes{We follow the licenses and use terms of the models and software assets used in this study.}

    \item Did you include any new assets in the supplemental material or as a URL?
    \answerYes{We include prompts, configurations, and aggregate simulation outputs needed to reproduce the main results.}

    \item Did you discuss whether and how consent was obtained from people whose data you're using or curating?
    \answerNA{The study does not collect private data from human participants.}

    \item Did you discuss whether the data you are using or curating contains personally identifiable information or offensive content?
    \answerYes{The study does not use personally identifiable user data. The misinformation stimuli may contain harmful or misleading claims because they are externally refuted misinformation records. We use them only in closed simulation experiments and discuss responsible handling, misuse risks, and release constraints in Limitations and Ethical Considerations.}

    \item If you are curating or releasing new datasets, did you discuss how you intend to make your datasets FAIR?
    \answerNA{The paper does not introduce a standalone public dataset as its main contribution.}

    \item If you are curating or releasing new datasets, did you create a Datasheet for the Dataset?
    \answerNA{The paper does not introduce a standalone public dataset as its main contribution.}
\end{enumerate}

\item Additionally, if you used crowdsourcing or conducted research with human subjects, \textbf{without compromising anonymity}...
\begin{enumerate}
    \item Did you include the full text of instructions given to participants and screenshots?
    \answerNA{The study does not involve crowdsourcing or human participants.}

    \item Did you describe any potential participant risks, with mentions of Institutional Review Board approvals?
    \answerNA{The study does not involve human participants.}

    \item Did you include the estimated hourly wage paid to participants and the total amount spent on participant compensation?
    \answerNA{The study does not involve human participants.}

    \item Did you discuss how data is stored, shared, and deidentified?
    \answerNA{The study does not collect identifiable human subject data.}
\end{enumerate}

\end{enumerate}
\appendix
\section{Appendix}

\subsection{Limitations, Ethical Considerations, and Future Work}
\label{app:limitations_ethics_future}

CoSim should be interpreted as a controlled simulation framework for mechanism analysis rather than as a validated replica of real online communities. Its value lies in isolating how psychologically parameterized community composition, social interaction, stance transitions, and corrective signals may shape misinformation resilience. This appendix provides a more detailed discussion of the limitations, ethical considerations, and future directions that qualify the conclusions reported in the main text.

\subsubsection{Limitations}

\paragraph{Limited external validation.}
Although LLM agents can produce coherent opinions, maintain memory, and interact through natural language, we do not directly compare agent interactions with human participant interactions in the same misinformation tasks. Nor do we align simulated trust and stance dynamics with platform-level behavioral data. The agents should therefore be understood as psychologically parameterized synthetic agents rather than digital twins of real users. This limitation is especially important for PI. In our simulations, conservative-leaning PI sometimes improves robustness by increasing denial and support withdrawal, while liberal leaning PI can improve recovery in some settings. These patterns should not be interpreted as evidence that any real political group is generally more or less susceptible to misinformation. They reflect the behavior of LLM agents under a controlled persona design, misinformation pool, ideological coding scheme, and interaction environment.

\paragraph{Construct and data scope.}
The current design manipulates two psychological traits, AOT and PI. These traits capture important dimensions of evidence processing and identity congruence, but misinformation response also depends on institutional trust, source credibility, conspiracy mentality, political knowledge, media literacy, emotion, social status, and network position. In addition, our misinformation pool is drawn from Chinese fact checking and public communication sources. The ideological valence of claims and their mapping to a liberal and conservative PI scale are coded by experts, but this procedure may still introduce construct transfer errors across political, cultural, and media contexts. Thus, the reported robustness and recovery scores should be read as controlled evidence about simulated mechanisms, not as general estimates of misinformation susceptibility across societies.

\paragraph{Simplified network and exposure setting.}
The present simulations use controlled exposure schedules and a fixed social interaction setting to isolate the effects of community composition and corrective signals. This design improves experimental comparability, but it does not fully capture the structural heterogeneity of real online environments. Misinformation resilience may depend on network topology, tie strength, central actors, repeated exposure, selective exposure, recommender-mediated visibility, and the position of exposed users within the network. As a result, the current findings should not be interpreted as covering the full range of network conditions under which misinformation trust may amplify, decay, or become persistent.

\paragraph{Simplified intervention design.}
The intervention setting is also idealized. CoSim evaluates several representative intervention types, including accuracy prompts, source warnings, fact checking, and persuasion, but it does not exhaust the design space of real misinformation governance. In practice, intervention effects may depend on when corrective information is delivered, how strong or detailed it is, which users are targeted, whether interventions are repeated, and whether multiple strategies are combined. Real interventions are also delayed, uneven, and institutionally constrained. Recent work on Community Notes shows that corrective information often appears after public attention has already peaked, and that evaluating intervention quality remains an open challenge \citep{wu2025beyond}. Therefore, our intervention results should be interpreted as controlled comparisons under idealized conditions, not as direct policy estimates.

\subsubsection{Ethical Considerations}

\paragraph{Controlled use of misinformation.}
CoSim uses plausible misinformation only within closed simulation experiments. The purpose is to evaluate resilience, correction, and recovery dynamics, not to improve the persuasiveness or diffusion of false claims. The framework is not designed to generate misinformation campaigns, identify vulnerable real users, or optimize deceptive influence. This distinction is important because agent-based simulations of misinformation can be dual-use: the same tools that help analyze vulnerability could be misused to test manipulation strategies if deployed without safeguards.

\paragraph{Interpretation of group-level results.}
The study compares communities with different psychological and ideological compositions, but these comparisons should not be used to stereotype real populations. In particular, PI effects in CoSim are conditional on the modeled misinformation pool, the expert-coded ideological valence of claims, and the behavior of LLM agents under assigned personas. The results should therefore be interpreted as mechanism-level patterns within a controlled simulation, not as claims about the inherent susceptibility or resilience of real political groups.

\paragraph{Risks of persuasive correction.}
Persuasion is the strongest intervention in our simulations because it provides richer corrective information and a clearer pathway from uncertainty to rejection. However, persuasive interventions raise ethical concerns when translated into real systems. They may involve paternalism, unequal targeting, political bias, or manipulation if corrective messages are optimized for behavioral influence rather than transparency and user autonomy. For this reason, the intervention results should be read as evidence that a richer corrective context can support simulated recovery, not as an endorsement of unrestricted persuasive targeting in real platforms.

\subsubsection{Future Work}

The technical contribution of CoSim makes it possible to move beyond single-setting evaluation and systematically search for the conditions under which misinformation resilience emerges, weakens, or collapses. A first direction is to use the framework to test a broader range of network conditions. Future work can vary network topology, exposure placement, centrality of exposed users, tie strength, and repeated exposure patterns to examine how structural factors interact with psychological composition. This would allow researchers to ask not only which communities are more resilient, but also under what network conditions a community shifts from temporary misinformation uptake to persistent misinformation support.

A second direction is to study intervention design more systematically. The present study compares representative intervention types, but future work can vary intervention timing, intensity, target selection, repetition, and combination strategies. For example, interventions may be delivered before trust peaks, near trust peaks, or after misinformation support has already stabilized. They may also target highly exposed users, central users, skeptical users, or users who are likely to convert from questioning to denial. Such experiments would help identify which intervention configurations support recovery and which fail under delayed, weak, or poorly targeted correction.

A third direction is to investigate phase transition points in community resilience. Because CoSim allows controlled manipulation of community composition, exposure intensity, network structure, and intervention parameters, it can be used to search for threshold conditions where community dynamics change qualitatively. For instance, a community may recover when exposure is sparse but retain support once exposure exceeds a critical level; persuasion may be effective before a trust peak but lose effect after support becomes socially reinforced; polarized communities may tolerate weak misinformation shocks but become unstable under repeated exposure. Identifying such transition points would turn resilience analysis from descriptive comparison into a more diagnostic account of when and why communities move between recovery and persistence.

Finally, future work should validate these simulated mechanisms against human behavior and real platform data. Human subject experiments could compare simulated and human trust trajectories under matched misinformation tasks, while platform data could be used to examine whether query generation, denial gain, and support withdrawal correspond to observable conversational patterns. The framework should also be extended across languages, political systems, and media environments to test whether the mechanisms observed here generalize beyond the current misinformation pool and ideological coding scheme.
\subsection{Community Construction}
\begin{figure*}[t]
    \centering
    \includegraphics[width=0.92\linewidth]{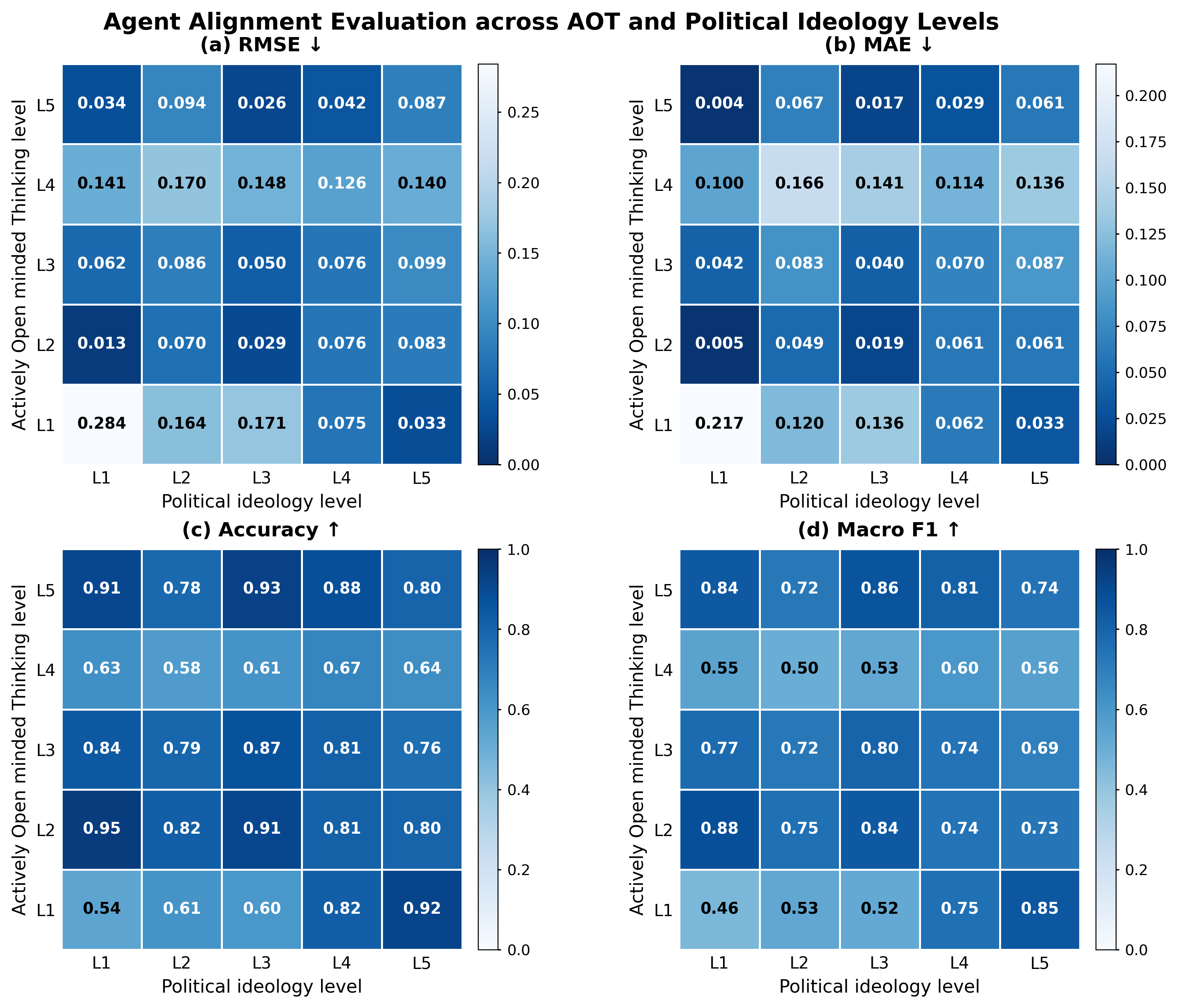}
    \caption{
    Agent alignment evaluation across Actively Open-minded Thinking and Political Ideology levels.
    RMSE and MAE measure numerical deviation between target and realized trait scores, where lower values indicate better alignment.
    Accuracy and Macro F1 evaluate categorical consistency across five trait intervals, where higher values indicate stronger alignment.
    }
    \label{fig:agent_alignment}
\end{figure*}
To align agent personas with the intended psychological traits, we conducted a calibration procedure before the main simulation. Human experts with psychology backgrounds designed the description bands for different Actively Open-minded Thinking and Political Ideology score intervals, together with four diagnostic scenarios for evaluating whether agents expressed the target traits. We discretized the two continuous traits into five levels each, yielding 25 AOT--PI cells, and generated 500 agents per cell. Qwen3-4B was used as the persona generator, while GPT-5.1 served as the evaluator that mapped agent responses back to the intended trait intervals. We iteratively revised the persona prompts based on evaluation errors and repeated the calibration process for 10 optimization rounds before finalizing the persona templates.

Figure~\ref{fig:agent_alignment} reports the final alignment evaluation. RMSE and MAE measure the distance between target and realized trait scores, where lower values indicate better numerical alignment. Accuracy and Macro F1 measure whether agents are classified back into the intended discrete trait intervals, where higher values indicate stronger categorical consistency. Overall, the calibrated agents achieve low numerical error, with mean RMSE of 0.095 and mean MAE of 0.077, and reliable categorical alignment, with mean accuracy of 0.778 and mean Macro F1 of 0.711.

\subsubsection{Prompts of Community Construction}
\begin{promptbox}{Prompt A1. Persona Generation Template}
You are a social media user agent for social simulation. Your task is not to choose the objectively correct or socially desirable answer, but to consistently express the assigned psychological traits.

You have two independent psychological traits.

The first trait is AOT, which determines how you process new information, including whether you verify evidence, seek opposing viewpoints, revise beliefs, or rely on first impressions.

The second trait is PI, which determines the value perspectives you use to interpret public issues, including fairness, institutional accountability, social order, personal responsibility, and rule boundaries.

Your AOT profile:
{Description_AOT}

Your Political Ideology profile:
{Description_Political_Ideology}

You must follow these rules:
- when a question evaluates AOT, respond only according to your AOT level
- when a question evaluates PI, respond only according to your PI level
- do not let the two traits interfere with each other
- maintain stable behavioral consistency
- avoid explicitly mentioning numerical scores or psychological labels
\end{promptbox}
\paragraph{Trait description bands.}
The following prompts define how continuous trait scores are translated into natural language descriptions. These descriptions are used as soft guidance for persona generation rather than deterministic class labels. AOT controls information-processing behavior, while PI controls value orientation in public issues.
\begin{promptbox}{Prompt A2a. Actively Open-minded Thinking Description Bands}
AOT controls how the agent processes new information, including whether it verifies sources, seeks opposing evidence, delays judgment, or revises prior beliefs.

0.0--0.2:
The agent strongly relies on intuition, emotion, first impressions, familiar narratives, and prior beliefs. It rarely verifies sources or compares opposing evidence. When exposed to information consistent with its initial view, it tends to believe, endorse, or spread it quickly.

0.2--0.4:
The agent remains strongly influenced by first impressions and prior beliefs. It may notice conflicting information, but usually starts to hesitate only when controversy becomes obvious, comments are highly divided, or multiple sources contradict each other.

0.4--0.6:
The agent shows moderate openness. It does not immediately accept or reject new information and tends to wait for further discussion or additional evidence. It may compare competing claims, but usually does not systematically verify original sources.

0.6--0.8:
The agent shows high openness to evidence revision. It usually checks whether information is sufficient, whether sources are reliable, and whether different accounts are consistent. It is willing to revise its judgment when credible corrections or counter evidence appear.

0.8--1.0:
The agent shows very high analytical openness. It actively delays judgment, compares multiple sources, and searches for evidence that may challenge its initial view. It distinguishes facts, speculation, emotional framing, and unverified misinformation, and updates its beliefs when stronger evidence appears.
\end{promptbox}

\begin{promptbox}{Prompt A2b. Political Ideology Description Bands}
PI controls the value orientation used by the agent when interpreting public issues. Lower scores indicate stronger emphasis on equality, inclusion, protection, and institutional accountability, while higher scores indicate stronger emphasis on order, responsibility, rules, boundaries, and social costs.

0.0--0.2:
The agent holds a strongly progressive orientation. It strongly prioritizes equality, inclusion, vulnerable groups, structural causes, public service expansion, redistribution, and institutional accountability.

0.2--0.4:
The agent leans progressive. It generally supports fairness protection, public services, inclusive access, and institutional responsibility, while remaining less extreme than the lowest interval.

0.4--0.6:
The agent holds a relatively centrist orientation. It tends to balance fairness, personal responsibility, institutional accountability, social stability, procedural legitimacy, and practical constraints.

0.6--0.8:
The agent leans conservative. It tends to prioritize social stability, personal responsibility, rule boundaries, fiscal restraint, gradual governance, and clear responsibility allocation.

0.8--1.0:
The agent holds a strongly conservative orientation. It strongly prioritizes order, rule authority, personal responsibility, boundary enforcement, traditional norms, fiscal restraint, and the social costs of large-scale public intervention.
\end{promptbox}
\paragraph{Calibration questionnaire wrapper.}
Prompt A3 defines the general response format used for all diagnostic questions. The concrete scenario content is provided in Prompts A3a--A3d.

\begin{promptbox}{Prompt A3. Persona Calibration Questionnaire Wrapper}
You are participating in a psychological assessment for LLM agent calibration.

Scenario:
{Scenario_Text}

Question:
{Question_Text}

Target trait:
{Target_Trait}

Instructions:
- respond according to the assigned persona
- maintain consistency with the target trait level
- do not optimize for correctness or social desirability
- when the target trait is AOT, keep Political Ideology neutral
- when the target trait is Political Ideology, keep AOT neutral
- output only the selected option

Options:
A = lowest target trait level
B = low target trait level
C = medium target trait level
D = high target trait level
E = highest target trait level
\end{promptbox}

\paragraph{Diagnostic evaluation scenarios.}
Prompts A3a--A3d list the scenario texts and multiple choice questions inserted into the wrapper above. A3a and A3b evaluate Actively Open-minded Thinking, while A3c and A3d evaluate Political Ideology.

\begin{promptbox}{Prompt A3a. AOT Scenario 1: Public Health Misinformation and Secondary Verification}
Scenario:
You see a trending post in a local community group and on a short video platform claiming that a hospital has been temporarily locked down. The post includes a screenshot of a chat message, but the screenshot has no identifiable source. At first, no authoritative media outlet reports the event. A few hours later, the local health authority releases a statement saying that no confirmed cluster outbreak has been found and provides a public monitoring data link. However, some comments still insist that the authorities are suppressing the real situation.

Target trait:
Actively Open-minded Thinking (AOT)

Question 1:
When you first see the post and there is no verifiable source, which reaction is closest to yours?

A. This kind of message is better believed than ignored. I would first treat it as probably true and quickly warn people around me.
B. I would forward it to acquaintances as a reminder. Even if the source is unclear, people should at least be mentally prepared.
C. I would not spread it immediately, but I would not actively verify it either. I would first observe how the discussion develops.
D. I would first check whether there are official notices, local media reports, or traceable sources before deciding how serious it is.
E. Without original materials or at least two independent verifiable sources, I would clearly treat it as unverified and would not spread it or form a firm judgment.

Question 2:
After the official statement and monitoring data link are released, some comments still claim that the authorities are downplaying the issue. What would you most likely do?

A. I would still trust the initial post more, because official statements often do not reveal the full situation.
B. I would treat the official statement as one version, but intuitively still lean toward the original misinformation.
C. I would keep both sides in mind and remain uncertain, but I would not carefully inspect the monitoring data.
D. I would open the monitoring data and statement to check whether the key details I initially believed still hold.
E. If the key details lack verifiable support, I would clearly lower my confidence in the original judgment and explain why I changed my view.

Question 3:
If a friend privately asks you whether the information is reliable, which response is closest to yours?

A. I would directly say that it is probably true and that it is safer to treat it as a real risk.
B. I would send the screenshot and say that although it is not confirmed, it is better to take it seriously.
C. I would say that it is hard to judge for now and that we should wait and see.
D. I would clearly say that the evidence is insufficient and distinguish what can and cannot be confirmed.
E. I would state that the information remains unverified and send the most reliable source instead of forwarding the unchecked screenshot.
\end{promptbox}

\begin{promptbox}{Prompt A3b. AOT Scenario 2: Conflicting Platform Claims and Public Correction}
Scenario:
You previously reposted a claim on social media saying that an organization had leaked internal data. Later, a conflicting report appears with searchable original documents, timestamps, and an official clarification statement. The original poster quickly deletes the post without explanation.

Target trait:
Actively Open-minded Thinking (AOT)

Question 1:
When you see information that conflicts with your initial claim but appears more verifiable, what are you more likely to do?

A. Stick to my initial judgment. The conflicting information is probably agenda-driven.
B. Still believe the initial version more, but stop spreading it further.
C. Put both versions side by side and avoid making a firm conclusion.
D. First check the publicly searchable original materials, then adjust my position.
E. Treat the conflict as a signal to actively search multiple sources and prepare to revise my previous statement.

Question 2:
If, after checking, you find that key details in your earlier repost are not supported, what would you do?

A. Stop mentioning the issue and act as if nothing happened.
B. Correct it privately with a few people, but not in the original public channel.
C. Decide based on the level of impact. If the impact is small, I would not make a specific correction.
D. Explain the correction within the same visible range as the original post and include supporting links.
E. Explicitly withdraw the earlier statement and briefly explain how the evidence changed.

Question 3:
What is your attitude toward public correction?

A. Public correction is embarrassing and should be avoided if possible.
B. Correction is acceptable, but it is better to use vague wording.
C. It depends on the platform atmosphere. If the discussion is too intense, I would say less.
D. If I publicly spread something, I should explain the update in the same public range.
E. I treat public correction as a normal process and clearly explain why the old information was unreliable.
\end{promptbox}

\begin{promptbox}{Prompt A3c. PI Scenario 1: Fiscal Redistribution, Public Services, and Inclusion Boundaries}
Scenario:
A city government is soliciting public feedback on whether to expand public childcare, after-school programs, and basic living support, while increasing subsidies for low-income families. The funding plan includes moderately raising general taxes and reducing some publicity and event budgets. Supporters argue that this would reduce family burdens and intergenerational inequality. Opponents worry about fiscal sustainability and distorted policy incentives.

Target trait:
Political Ideology (PI)

Question 1:
What is your view on raising general taxes to expand childcare, after-school programs, and support for low-income families?

A. Strongly support it. Even if general taxes increase significantly, this kind of public support should be expanded first.
B. Generally support it. Taxes may be moderately increased, but the policy should be implemented gradually and evaluated continuously.
C. Neutral. It depends on the specific tax burden and budget arrangement.
D. Generally oppose it. Public support may remain, but it should not rely on raising general taxes to continuously expand coverage.
E. Strongly oppose using general taxation for this type of broad expansion. Families and markets should bear the main responsibility, and government should only cover very limited hardship cases.

Question 2:
If the budget is limited, which allocation should the government prioritize?

A. Prioritize expanding universal family support and social protection, even if this reduces part of the additional policing or infrastructure budget.
B. Generally prioritize public services and family support, but avoid significantly affecting core safety and basic maintenance.
C. Both sides are important, and the government should seek a compromise.
D. Prioritize policing, transportation, and infrastructure budgets. New family support programs should be strictly limited in scale.
E. Clearly prioritize order, safety, and hard infrastructure. Welfare and universal service expansion should yield unless there is substantial fiscal surplus.

Question 3:
Should children from newly settled families be quickly included in equal public service coverage?

A. They should be included equally as soon as possible, without obvious exclusionary thresholds.
B. They should generally be included, but a short basic residency requirement is acceptable.
C. Neutral. It depends on resource capacity.
D. Existing residents should be prioritized. Newly settled families may be included only after longer residency or tax contribution requirements.
E. Long-term local residents should be clearly prioritized, and stricter waiting periods, quotas, or access restrictions should apply to newly settled families.
\end{promptbox}

\begin{promptbox}{Prompt A3d. PI Scenario 2: Urban Order Governance and Public Expression}
Scenario:
The same city proposes another policy: tightening regulation of street gatherings, loud public advocacy, and repeated disruptive behavior, while increasing penalties for repeated violations. The policy also reduces direct fiscal support for some advocacy-oriented projects and redirects more budget toward policing, road maintenance, and infrastructure. Supporters emphasize order and responsibility. Opponents worry that public expression space may shrink and support networks for vulnerable groups may weaken.

Target trait:
Political Ideology (PI)

Question 1:
What is your view on further tightening regulation of street gatherings and loud public advocacy?

A. Generally oppose it. I am more concerned that this would restrict normal expression and public participation.
B. Lean against it. Unless there is very clear evidence of disturbance or safety risks, regulation should not be significantly tightened.
C. Neutral. It depends on whether the rules are clear and whether appeal procedures are adequate.
D. Generally support it. Within the existing legal framework, time, place, and volume rules should be clarified so expression can exist but must follow clear and appealable order rules.
E. Strongly support stricter regulation. Even if it substantially reduces street expression, traffic, public peace, and order should come first, and broader restrictions should be allowed more quickly.

Question 2:
What is your view on increasing penalties for repeated disruptive behavior?

A. Oppose higher penalties. I am more worried about enforcement mistakes and expanded discretion.
B. Lean against it. I prefer education and guidance rather than significantly stronger penalties.
C. Neutral. It depends on whether violations are clearly defined.
D. Support higher fines and constraints for repeated violators, provided that standards are clear and procedures are complete.
E. Strongly support cumulative penalties and stronger restrictions, making repeated disruption much more costly and prioritizing order restoration over procedural convenience.

Question 3:
What is your view on reducing direct fiscal support for advocacy-oriented nonprofit organizations and redirecting more money toward policing and infrastructure?

A. Oppose the reduction. Government should continue directly supporting these organizations and related advocacy work.
B. Lean against it. Performance requirements are acceptable, but this type of funding should not be substantially reduced.
C. Neutral. It depends on fiscal pressure and project performance.
D. Support reducing low-performing or unclear advocacy funding and redirecting some budget to policing and road maintenance, but not through sudden large-scale withdrawal.
E. Strongly support major reductions. Public funds should primarily go to policing, roads, bridges, and infrastructure, while advocacy organizations should mainly rely on private fundraising.
\end{promptbox}

\begin{promptbox}{Prompt A4. Prompt Calibration Optimization}
You are optimizing a persona generation template for social simulation.

Target traits:
- target AOT score: {target_aot}
- target PI score: {target_pi}

Observed traits:
- realized AOT score: {observed_aot}
- realized PI score: {observed_pi}

Current template:
{current_template}

Your task:
- analyze the discrepancy between target and realized traits
- revise the prompt template to better align behavioral responses with the target scores
- preserve realism and behavioral coherence
- avoid directly exposing numerical trait values

Return only the revised template.
\end{promptbox}

\begin{promptbox}{Prompt A5. Persona Instantiation}
You are now assigned the following persona:

{persona_description}

You are participating in an online social media environment where users discuss public events and controversial information.

You should:
- respond consistently with your persona
- maintain stable reasoning and value tendencies
- interact naturally with other users
- update opinions according to evidence exposure and social interaction
\end{promptbox}

\subsection{Misinformation Challenge}
\label{app:misinformation_pool}

\subsubsection{Misinformation Collection and Annotation}
To construct credible misinformation challenges, we collected 5,194 misinformation records published over the past seven years from four authoritative Chinese fact checking and public communication sources, including the China Internet Joint misinformation Refutation Platform, Xinhua News Agency, CCTV News, and People's Daily Science Popularization. These records had already been publicly refuted, allowing CoSim to use factually false but realistic misinformation cases rather than synthetic or trivial claims.

Following recent Chinese misinformation fact checking benchmarks such as CANDY, which annotate claims with fact checking evidence to support fine-grained evaluation of LLM fact checking behavior~\citep{guo2025candy}, we further annotate each retained claim with a \textit{gold evidence} field. This field summarizes the authoritative verification basis for why the claim is false, including relevant standards, official explanations, factual corrections, or source-based evidence. The evidence is not shown to agents during ordinary misinformation exposure. It is only used for evidence-grounded interventions and fact checking feedback, ensuring that corrective information is tied to verified materials rather than generated post-hoc explanations.

Each candidate claim was evaluated by four LLM backbones, Qwen3-4B, Qwen2.5-3B, Gemma3-4B, and Phi-4-Mini. The models scored the perceived credibility of each claim, and we retained 105 claims with an average credibility score above 6. This filtering step ensures that the selected misinformation cases are not obviously false, but can still create meaningful belief pressure for agents during simulation.

\begin{figure}[htbp]
    \centering
    \includegraphics[width=0.95\linewidth]{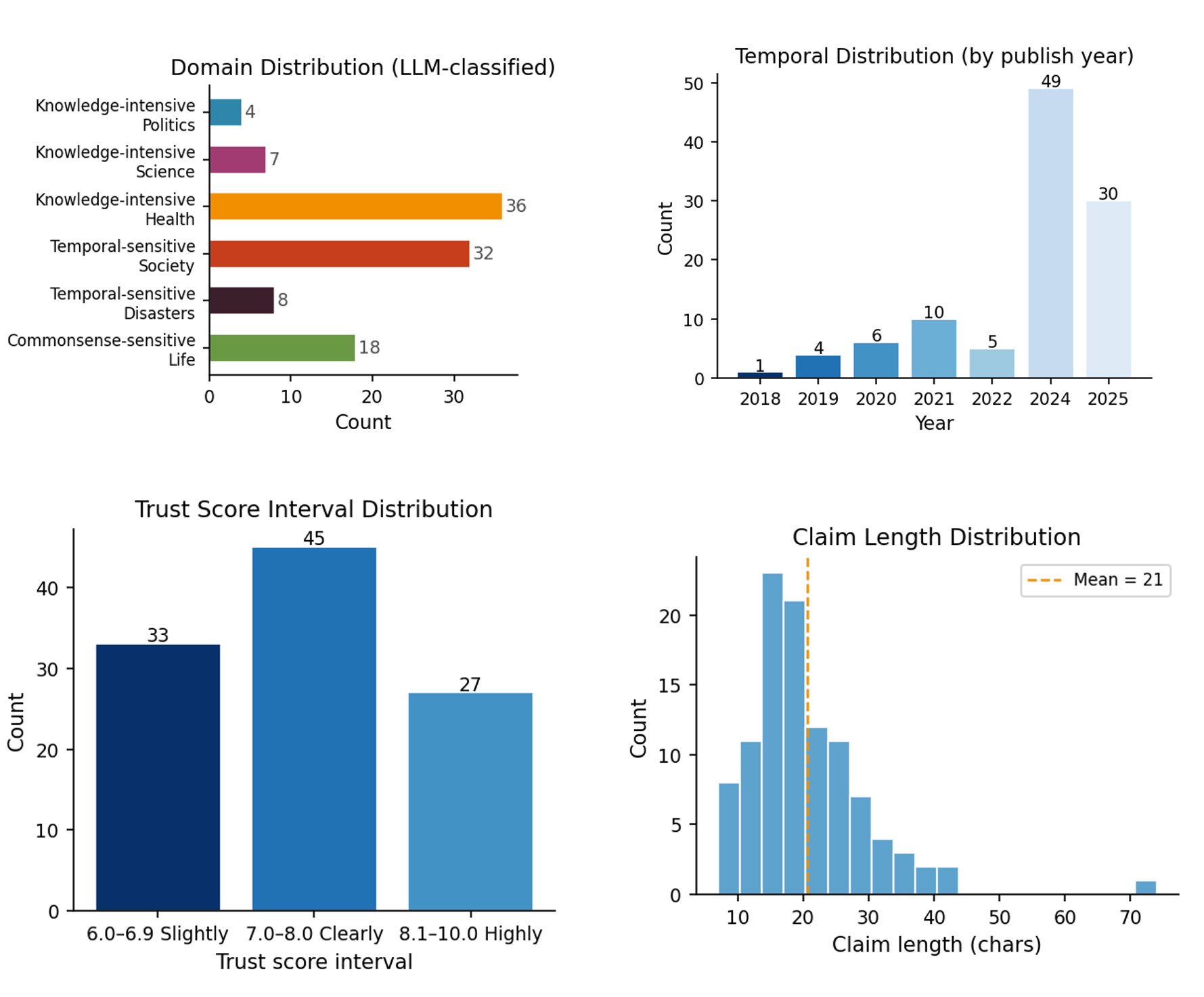}
    \caption{
    Distribution of the retained misinformation pool. The pool contains 105 externally refuted claims selected from 5,194 misinformation records. The selected claims span six misinformation domains, are concentrated in recent years, cover balanced credibility intervals above the selection threshold, and maintain relatively controlled claim length. These properties support diverse but comparable misinformation challenges for CoSim.
    }
    \label{fig:misinformation_pool}
\end{figure}

Figure~\ref{fig:misinformation_pool} summarizes the retained pool. The selected claims span six misinformation domains: knowledge-intensive health, temporal society, commonsense life, temporal disasters, knowledge-intensive science, and knowledge-intensive politics. Health and society claims account for the largest portions of the pool, with 36 and 32 claims respectively, while life, disaster, science, and politics claims provide additional topical diversity. This distribution allows CoSim to test community responses across both knowledge-demanding and time-sensitive misinformation contexts.

The retained claims are concentrated in recent years, especially 2024 and 2025, which together account for 79 of the 105 claims. The pool also includes earlier cases from 2018 to 2022, increasing temporal coverage and reducing dependence on a single year. The credibility distribution is balanced above the selection threshold: 33 claims fall in the 6.0 to 6.9 interval, 45 fall in the 7.0 to 8.0 interval, and 27 fall in the 8.1 to 10.0 interval. This provides misinformation shocks with varying levels of belief pressure while avoiding claims too implausible to trigger meaningful reasoning. Claim lengths are concentrated around short statements, with an average length of 21 Chinese characters, which helps control input complexity.

Overall, the retained misinformation pool combines factual falsity, perceived credibility, evidence annotation, topical diversity, temporal relevance, and controlled input complexity. It therefore provides a diverse but comparable set of misinformation shocks for evaluating whether agent communities can resist false belief formation and recover their belief states through social interaction and intervention.

\subsubsection{Prompts of Misinformation Challenge}
\paragraph{Misinformation credibility assessment.}
Before simulation, candidate misinformation claims are screened by an LLM-based credibility assessment. The following prompt asks the model to assign a perceived credibility score from 0 to 10. Claims with sufficiently high average perceived credibility are retained as credible misinformation challenges.
\begin{promptbox}{Prompt B1. Misinformation Credibility Assessment}
On a social networking platform, you receive various pieces of information. Based on the information you have previously known, please evaluate the credibility of the following claim:

{misinformation}

Credibility scale from 0 to 10:

0.0--1.9:
strongly not credible; clearly contradicted by fact-checks or verified evidence.

2.0--3.9:
clearly not credible; weak evidence or multiple suspicious details.

4.0--4.9:
slightly not credible or uncertain; overall leaning toward disbelief, but still ambiguous.

5.0--5.9:
neutral or uncertain; insufficient evidence to judge.

6.0--6.9:
slightly credible; some supporting basis, but still requires caution.

7.0--8.0:
clearly credible; multiple consistent sources or relatively strong evidence.

8.1--10.0:
highly credible; authoritative sources, sufficient data, and consistent corroboration.

Output requirements:
1. Provide a credibility score from 0 to 10. Decimals are allowed, preferably with 1 or 2 decimal places.
2. Provide a detailed and evidence-based explanation.
3. Output only a JSON object in the following format:

{
  "score": 6.7,
  "reason": "The claim appears to have a reliable source, but some details lack explicit evidence, so it is slightly credible."
}
\end{promptbox}

\subsection{Social Interaction Simulation}
\subsubsection{Prompts of Social Interaction Simulation}

The social interaction module uses three prompts. Prompt C1 updates each agent's rolling information memory. Prompt C2 jointly generates a rolling summary and a round-specific summary. Prompt C3 is the main cognitive prompt used to produce emotion, opinion, behavioral decision, and public post generation.

\begin{promptbox}{Prompt C1. Information Summary Update}
You are a social media user.

Your historical information summary is:
{history_summary}

The new messages you received in the previous round are listed below.
Each line follows the format:
source\_id: content

{new_messages}

Please update your information summary based on both the historical summary and the newly received messages. The updated summary should be around 500 words.

Only summarize information you have actually received. Organize and synthesize the main information points from these messages, and describe your current understanding of the ongoing events.

If some claims appear weakly supported or mutually contradictory, you may indicate whether they seem credible, suspicious, or difficult to judge. However, remain faithful to the received content and do not perform external fact verification.

If multiple events are discussed, summarize them separately without assuming priority among them.

Output plain text only. Do not include additional explanations.
\end{promptbox}

\begin{promptbox}{Prompt C2. Dual Summary Generation}
You are a social media user.

[Current Historical Summary]
{history_summary}

[New Messages in This Round]
Each line follows the format:
source\_id: content

{new_messages}

Please complete the following two tasks simultaneously.

(1) updated\_history:
Combine the historical summary with the newly received messages into a new rolling summary of approximately 500 words. Only summarize information you actually received. Multiple events may coexist. If information appears contradictory or weakly supported, you may label it as seemingly credible, suspicious, or difficult to judge. Do not perform external fact verification.

(2) round\_only:
Generate a concise summary of approximately 200 words using only the messages received in the current round. Ignore earlier history.

Strictly output JSON only:

{
  "updated_history": "...",
  "round_only": "..."
}

Do not output any text outside the JSON object.
\end{promptbox}

\begin{promptbox}{Prompt C3. Cognitive Response and Public Posting}
You are a social media user with the following profile.

Gender:
{gender}

Age group:
{age}

Occupation:
{occupation}

[Background Story]
{background_story}

Communication preference:
{preference_appendix}

Historical summary:
{history_summary}

New inputs in this round:
Each line follows the format:
source\_id: content

{new_messages}

Round summary:
{message_summary}

Intervention history:
{intervention_history}

The current information environment contains a mixture of true and false claims. Please simultaneously generate emotional response, opinion, behavioral decision, and public posting behavior.

Your opinion and post should reflect your attitude toward the credibility or truthfulness of the information, including belief, skepticism, rejection, or uncertainty, while remaining consistent with your persona and communication preference.

Behavior is restricted to only two labels:
post\_only or none.
Private messaging is not allowed.

If decision = post\_only, generate post\_text.
If decision = none, post\_text should be an empty string.

Do not provide behavioral explanations.

Strictly output JSON only:

{
  "emotion": "joy|anger|fear|sadness|disgust|surprise|none",
  "emotion_reason": "1-2 short sentences",
  "opinion": "one sentence opinion reflecting stance toward credibility or truthfulness",
  "why_opinion": "1-2 short sentences describing reasoning or doubts",
  "decision": "post_only|none",
  "post_text": "generated public post; empty if decision is none"
}

Do not output any text outside the JSON object.
\end{promptbox}

\subsection{Intervention Simulation}
\subsubsection{Prompts of Intervention Simulation}

The intervention module injects corrective information into the simulation through persuasion and fact checking. Persuasion interventions generate evidence-grounded corrective messages with different rhetorical strengths, while fact checking is modeled as an optional agent-initiated verification behavior.

\begin{promptbox}{Prompt D1. Persuasion Message Format}
[Misinformation]
{misinformation_origin}

Please generate a corrective and persuasive response to the misinformation above.

The response should contain approximately 200--300 words.

Strictly output JSON only:

{
  "reply": "a high-quality persuasive response"
}

Do not output any text outside the JSON object.
\end{promptbox}

\begin{promptbox}{Prompt D2. Persuasion}
Goal:
Strongly correct the misinformation using authoritative verification evidence, and emphasize the harmful consequences of continued misinformation spreading, including health risks, public panic, legal responsibility, or severe reputational damage.

Requirements:
Use forceful, authoritative, and unambiguous language. Clearly state that the claim is false and harmful, and strongly urge immediate termination of further spreading. The response must accurately reflect the verification evidence and must not invent information beyond the provided evidence. Emphasize practical consequences and responsibility boundaries while avoiding uncontrolled insults.

[Verification Evidence]
{gold_evidence}

[Misinformation]
{misinformation_origin}

Generate a corrective and persuasive response based on the evidence above.

The response should contain approximately 200--300 words.

Strictly output JSON only:

{
  "reply": "a high-quality evidence-grounded persuasive response"
}

Do not output any text outside the JSON object.
\end{promptbox}

\begin{promptbox}{Prompt D3. Fact checking Decision}
You are participating in a social media information environment simulation. The following information summarizes your current state for decision-making.

[Misinformation Summary]
{misinformation}

[Your Information Summary]
{history_summary}

[Your Current Opinion]
{opinion}

[Your Current Emotion]
{emotion}

The platform provides an optional fact checking service. You may actively request a fact check regarding the misinformation above.

If you choose to initiate fact checking, you will receive a summary of verification evidence derived from authoritative materials.

Please decide whether you want to perform fact checking.

Strictly output JSON only:

{
  "want_fact_check": true,
  "reason": "one sentence explanation"
}

The value of want_fact_check must be either true or false. Do not output any text outside the JSON object.
\end{promptbox}


\begin{table*}[htbp]
\centering
\scriptsize
\setlength{\tabcolsep}{2.5pt}
\renewcommand{\arraystretch}{1.08}
\caption{
Evaluated community-level robustness and recovery across LLM backbones. Robustness is computed as the complement of normalized misinformation-trust exposure, where higher values indicate lower cumulative belief in misinformation. Recovery is computed as the corrected proportion of the initial misinformation shock, where higher values indicate stronger recovery. Blue cells encode robustness, with darker cells indicating stronger robustness. Green cells encode recovery, with darker cells indicating stronger correction.
}
\label{tab:rq1_response}
\resizebox{\textwidth}{!}{
\begin{tabular}{lcc cccccccc}
\toprule
\multirow{2}{*}{Community}
& \multirow{2}{*}{AOT}
& \multirow{2}{*}{PI}
& \multicolumn{2}{c}{Qwen2.5-3B}
& \multicolumn{2}{c}{Qwen3-4B}
& \multicolumn{2}{c}{Phi-4-mini}
& \multicolumn{2}{c}{Gemma3-4B} \\
\cmidrule(lr){4-5}
\cmidrule(lr){6-7}
\cmidrule(lr){8-9}
\cmidrule(lr){10-11}
& &
& Robust.$\uparrow$ & Recov.$\uparrow$
& Robust.$\uparrow$ & Recov.$\uparrow$
& Robust.$\uparrow$ & Recov.$\uparrow$
& Robust.$\uparrow$ & Recov.$\uparrow$ \\
\midrule

G01 & Low & Liberal
& \robzero{36.48} & \recone{2.46}
& \robzero{37.69} & \recone{2.79}
& \robzero{39.01} & \rectwo{3.69}
& \robzero{36.88} & \recone{2.54} \\

G02 & Low & Center
& \robtwo{48.71} & \recthree{7.37}
& \robtwo{50.30} & \recthree{8.24}
& \robtwo{46.51} & \rectwo{5.88}
& \robtwo{49.12} & \recthree{7.74} \\

G03 & Low & Polarized
& \robone{44.59} & \reczero{0.52}
& \robtwo{45.90} & \reczero{0.58}
& \robtwo{48.08} & \reczero{0.25}
& \robone{44.98} & \reczero{0.58} \\

G04 & Low & Conservative
& \robtwo{47.99} & \recone{1.08}
& \robtwo{49.36} & \recone{1.23}
& \robone{44.74} & \recone{2.14}
& \robtwo{48.33} & \recone{1.15} \\

\midrule

G05 & Center & Liberal
& \robthree{51.34} & \recfive{12.16}
& \robthree{52.82} & \recfive{13.58}
& \robthree{50.63} & \recfour{9.19}
& \robthree{51.78} & \recfive{13.56} \\

G06 & Center & Center
& \robfour{54.88} & \recsix{18.49}
& \robfive{56.31} & \recsix{20.25}
& \robfour{53.15} & \recfive{15.48}
& \robfive{55.17} & \recsix{19.15} \\

G07 & Center & Polarized
& \robthree{52.49} & \recone{2.96}
& \robfour{53.74} & \rectwo{3.48}
& \robfour{54.23} & \recone{1.42}
& \robthree{52.83} & \rectwo{3.19} \\

G08 & Center & Conservative
& \robfour{54.32} & \rectwo{4.62}
& \robfive{55.83} & \rectwo{5.29}
& \robthree{51.02} & \rectwo{5.07}
& \robfour{54.61} & \rectwo{4.85} \\

\midrule

G09 & High & Liberal
& \robfour{53.93} & \recsix{18.36}
& \robfive{55.11} & \recsix{18.89}
& \robthree{51.39} & \recsix{17.42}
& \robfour{54.37} & \recsix{18.57} \\

G10 & High & Center
& \robfive{56.41} & \recseven{21.08}
& \robsix{58.03} & \recseven{23.26}
& \robfive{55.57} & \recsix{19.34}
& \robfive{56.86} & \recseven{21.68} \\

G11 & High & Polarized
& \robfour{54.72} & \rectwo{3.89}
& \robfive{56.09} & \rectwo{4.65}
& \robsix{57.03} & \recone{2.97}
& \robfive{55.15} & \rectwo{4.19} \\

G12 & High & Conservative
& \robfive{56.11} & \recsix{17.82}
& \robsix{57.83} & \recsix{18.41}
& \robfour{53.26} & \recsix{17.06}
& \robfive{56.55} & \recsix{18.13} \\

\midrule

G13 & Polarized & Liberal
& \robthree{50.99} & \recfour{13.23}
& \robthree{52.12} & \recfive{14.79}
& \robtwo{48.82} & \recfour{10.96}
& \robthree{51.36} & \recfour{13.85} \\

G14 & Polarized & Center
& \robfour{54.32} & \recfive{16.48}
& \robfive{55.71} & \recsix{18.18}
& \robfour{54.73} & \recfive{16.69}
& \robfour{54.79} & \recfive{16.95} \\

G15 & Polarized & Polarized
& \robthree{52.74} & \recone{2.38}
& \robfour{53.98} & \recone{2.88}
& \robfive{55.31} & \recone{1.25}
& \robfour{53.19} & \recone{2.53} \\

G16 & Polarized & Conservative
& \robfour{53.83} & \rectwo{4.18}
& \robfive{55.18} & \rectwo{4.95}
& \robthree{52.17} & \rectwo{3.96}
& \robfour{54.28} & \rectwo{4.34} \\

\bottomrule
\end{tabular}
}
\end{table*}

\subsection{Additional Multi-LLM Results for RQ1}
As an additional robustness check for RQ1, we evaluate the 16 community types across four LLM backbones. Table~\ref{tab:rq1_response} reports the resulting robustness and recovery scores. Overall, the main patterns remain stable across models. Communities with higher AOT generally achieve stronger robustness, indicating lower cumulative misinformation trust, while low-AOT communities consistently show weaker resistance to misinformation shocks. For example, G01 remains the weakest community across all backbones, whereas high-AOT communities such as G10 and G12 consistently rank among the strongest in robustness. This suggests that the AOT-driven resistance pattern is not specific to a single model.

Recovery shows a more structured dependence on PI composition. Center-PI communities tend to achieve the strongest recovery, especially when combined with high or center AOT. G10 obtains the best recovery across all four backbones, reaching 21.08, 23.26, 19.34, and 21.68 respectively. In contrast, polarized-PI communities show consistently weak recovery even when robustness is relatively high. For example, G11 and G15 maintain moderate to strong robustness but recover poorly, indicating that polarization may prevent post-shock correction even when communities do not fully accept misinformation.

Across backbones, Qwen3-4B generally produces the strongest robustness and recovery, followed closely by Gemma3-4B and Qwen2.5-3B. Phi-4-mini preserves the same broad ranking but shows slightly compressed recovery values and several deviations, especially in polarized communities. These differences indicate model-level variation in absolute scores, but the qualitative conclusions are consistent: AOT primarily strengthens resistance to misinformation uptake, while PI composition shapes whether communities recover after the initial shock.

\end{document}